\documentclass[twocolumn,floatfix,10pt,superscriptaddress,prx,longbibliography]{revtex4-1}
\usepackage[utf8]{inputenc}

\usepackage{natbib}
\usepackage{graphicx}
\usepackage{latexsym}
\usepackage{amssymb}
\usepackage{amsmath}
\usepackage{amsfonts}
\usepackage{gensymb}
\usepackage{bm}
\usepackage{hyperref}
\usepackage{braket}
\usepackage{physics}
\usepackage{bbold}
\usepackage[caption=false]{subfig}
\usepackage[dvipsnames]{xcolor}
\usepackage[normalem]{ulem}
\usepackage{siunitx}
\usepackage[version=4]{mhchem}

\DeclareRobustCommand{\rchi}{{\mathpalette\irchi\relax}}
\newcommand{\irchi}[2]{\raisebox{\depth}{$#1\chi$}}

\hypersetup{
  pdfauthor = {Taige Wang, Noah F. Q. Yuan, Liang Fu},
  pdftitle = {Wang et al. - 2020 - Moir\'e surface states and enhanced superconductivity in topological insulators}
}

\begin{document}

\title{Moir\'e surface states and enhanced superconductivity in topological insulators}

\author{Taige Wang}
\affiliation{Department of Physics, Massachusetts Institute of Technology, Cambridge, MA 02139, USA}
\affiliation{Department of Physics, University of California, San Diego, CA 92093, USA}
\affiliation{Department of Physics, University of California, Berkeley, CA 94720, USA}

\author{Noah F. Q. Yuan}
\thanks{Current address: Shenzhen JL Computational Science and Applied Research Institute, Shenzhen, 518109 China.}
\author{Liang Fu}
\affiliation{Department of Physics, Massachusetts Institute of Technology, Cambridge, MA 02139, USA}

\date{\today}

\begin{abstract}

Recently, moir\'e superlattices have been found on the surface of topological insulators (TI) due to the rotational misalignment of topmost layers. In this work, we study the effects of moir\'e superlattices on the topological surface states using a continuum model of Dirac electrons moving in a periodic potential. Unlike twisted bilayer graphene, moir\'e surface states cannot host isolated bands due to their topological nature. Instead, we find (high-order) van Hove singularities (VHS) in the moir\'e band structure that give rise to divergent density of states (DOS) and enhance interaction effects. Due to spin-momentum locking in moir\'e surface states, possible interaction channels are limited. In the presence of phonon mediated attraction, superconductivity is strongly enhanced by the power-law divergent DOS at high-order VHS. The transition temperature $T_c$ exhibits a \textit{power-law} dependence on the retarded electron-phonon interaction strength $\lambda^*$. This enhancement is found to be robust under various perturbations from the high-order VHS.

\end{abstract}


\maketitle

\section{Introduction}

In recent years, moir\'e superlattices have been studied extensively in various 2D van der Waals heterostructures exemplified by graphene and transition metal dichalcogenide (TMD) multilayers \cite{MacDonaldReview,AndreaReview,EfthimiosReview}. These moir\'e systems exhibit a variety of remarkable electronic properties due to strong correlation effects in flat minibands. Besides graphene and TMD, another large family of moir\'e superlattices can be found in topological insulators \cite{TangNaturePhysics,Bi2Se3Moire,Bi2Te3Moire,Bi2Se3/MoS2DFT,Bi2Se3/MoS2,TMD/Bi2Se3,graphene/Bi2Se3,FeSe/Bi2Se3,Au/Bi2Se3,hBN/Bi2Se3,NbSe/Bi2Se3}. When Bi$_2$Se$_3$ and Bi$_2$Te$_3$ bulk crystals are grown by the molecular-beam epitaxy (MBE), it is common to find a small rotational misalignment of topmost quintuple layers, leading to a moir\'e superlattice on the surface \cite{Bi2Se3Moire,Bi2Te3Moire}. Interestingly, a scanning tunneling microscope (STM) measurement \cite{Bi2Te3Moire} has directly observed such moir\'e superlattice in \ce{Bi2Te3} and found multiple sharp peaks in the local density of states (LDOS). Moreover, moir\'e superlattices can also be found in the van der Waals heterostructure of topological insulators and large-gap insulators \cite{Bi2Se3/MoS2DFT,Bi2Se3/MoS2,TMD/Bi2Se3,hBN/Bi2Se3}. Despite the ubiquity of moir\'e superlattices in TI, their effects on topological surface states have not been studied theoretically.

In this letter, we study moir\'e surface states of TI. The topological nature of TI surface states prevents them from gap opening as long as time-reversal symmetry is preserved, hence the moir\'e surface states do not form isolated mini bands, unlike other moir\'e systems such as graphene and TMD. Instead, we find prominent van Hove singularities (VHS) in moir\'e surface states which give rise to divergent density of states (DOS). Under appropriate conditions, some of these VHS exhibit power-law divergent DOS, which are known as \textit{high-order} VHS \cite{highVHS}.

We further study interaction effects near (high-order) VHS enhanced by the divergent DOS  \cite{Chubukov,Chubukov2,Supermetal,Guinea}. In moir\'e surface states, possible interaction channels at VHS are limited due to spin-momentum locking. Under attractive interactions, superconductivity is favored. We find a new analytic formula for the electron-phonon superconducting critical temperature $T_c$ (see Eq.~(\ref{eq:tc})), which exhibits a \textit{power-law} dependence of the retarded electron-phonon interaction $\lambda^*$ and is thus parametrically enhanced with respect to the exponentially small $T_c$ in ordinary metals and at ordinary VHS \cite{VHSSC,VHSSC2}. Importantly, the absence of moire band gaps and the large electron velocity away from VHS facilitate the reduction of Coulomb repulsion through retardation effects. We also show that the enhancement of superconductivity is robust and persist even when the system is perturbed away from high-order VHS.

This work is organized as follows: we first introduce and study a model of moir\'e surface states as Dirac fermion in a periodic scalar potential in Sec.~\ref{sec:Dirac}. Within the model, we identify high-order VHS at the crosses of circular Fermi surfaces. Then, we solve the gap equation for the superconducting critical temperature $T_c$ in the presence of power-law divergent density of states, taking account of both electron-phonon interaction and Coulomb repulsion within the Anderson-Morel approximation \cite{AndersonMorel} (Sec.~\ref{sec:SC}). In the end, we discuss several experimental platforms to search for moir\'e surface states and enhanced superconductivity.

\begin{figure*}[t]
    \centering
    \includegraphics[width=0.8\textwidth]{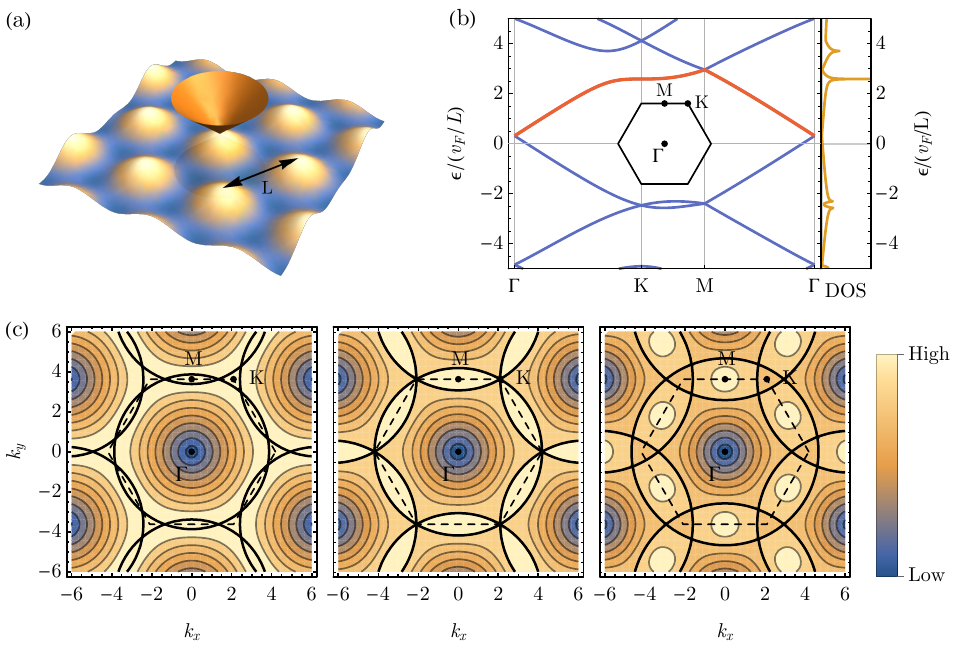}
    \caption{(a) Dirac fermion in a $C_3$ periodic potential with superlattice constant $L$. (b) Left panel: spectrum at potential $U_c = 1.36 v_F/L$. The corresponding mini BZ is shown in the center. The entire spectrum remains gapless due to the symmetry anomaly. Between the main Dirac point $\Gamma$ and satellite Dirac points $M$ (highlighted in red), we find a high-order van Hove singularity (VHS) at the $K$ point. Right panel: corresponding density of state (DOS). The high-order VHS peak stands out, with several ordinary VHS accompanied. (c) Fermi surface at potential $U = 1.0 v_F/L$, $1.36 v_F/L$ and $1.7 v_F/L$ from left to right. The ones passing through VHS are all perfect circles, which are plotted in thick black curves. When $U = 1.36 v_F/L$, all three Fermi surfaces cross at the $K$ point, making it a high-order VHS.}
    \label{fig:dirac}
\end{figure*}

\section{Dirac fermion in a periodic scalar potential} \label{sec:Dirac}

In this section, we introduce and study a model of moir\'e topological insulator surface states as Dirac fermion in a periodic scalar potential. A previous DFT study in \ce{Bi2Se3}/\ce{MoS2} \cite{Bi2Se3/MoS2DFT} revealed folded Dirac cones within the bulk gap due to the moir\'e superlattice. Thus, we start with the massless Dirac fermion in two dimensions (2D)
\begin{equation}\label{eq:dirac0}
    H_0(\bm{k}) = v_F (k_x \sigma^y - k_y \sigma^x),
\end{equation}
where $v_F$ is the Fermi velocity, $\bm{k} = (k_x, k_y)$ is the two-dimensional momentum, $\boldsymbol{\sigma} = (\sigma^x, \sigma^y)$ are the Pauli matrices. Now we allow the continuous translation symmetry be broken into discrete ones by the moir\'e superlattice, while leave the time-reversal symmetry intact. Then the lowest order perturbation can be described by a spin-independent periodic scalar potential $U(\bm r)$:
\begin{equation}\label{eq:dirac}
    H(\bm{k},\bm{r}) = H_0(\bm{k}) + U(\bm{r}) \sigma^0,
\end{equation}
where $\sigma^0$ is the identity matrix, $U(\bm{r}) = U(\bm{r} + \bm{L}_{1,2})$, and the $\bm{L}_{1,2}$ are two primitive vectors of the moir\'e superlattice. A schematic diagram of this setup is shown in Fig.~\ref{fig:dirac}a. This model can apply to bulk TI crystals with top layers twisted or the surface state in the interface between a topological insulator (TI) and a large-gap insulator.

The density of states (DOS) of the system described by Eq.~(\ref{eq:dirac}) generally looks like the right panel in Fig.~\ref{fig:dirac}b, where positive and negative sides are qualitatively similar.
Near zero energy, DOS grows linearly. As energy increases, VHS peaks emerge. At higher energy, new Dirac points are formed (known as satellite Dirac points \cite{Graphene}), so that the entire spectrum remains gapless. Among these VHS peaks, there are a few prominent ones that are, in fact, high-order VHS peaks, given appropriate parameters. 

In comparison to ordinary VHS that are caused by saddle points in the energy dispersion, such high-order VHS peaks are caused by high-order critical points $\bm k_c$ in momentum space, where the electron velocity and the Hessian matrix determinant both vanish, $\nabla E(\bm k_c)=\bm 0,\quad\det D(\bm k_c)=0,\quad (D_{ij}=\partial_i\partial_j E)$. Around these high-order VHS, the energy dispersion is characterized by high-order polynomials of momenta, and the DOS shows power-law divergence \cite{highCP},
\begin{equation} \label{eq:cpm}
    N(\xi)=
    \begin{cases}
    C_{+}\xi^\nu, & \xi>0\\
    C_{-}(-\xi)^\nu, & \xi<0
    \end{cases}.
\end{equation}
Here $-1<\nu<0$ is the power-law exponent and $C_{\pm}>0$ are coefficients of the electron (hole) side. $E(\bm k)$ denotes the energy dispersion and $\xi=E-E(\bm k_c)$.

To be concrete, we first consider a periodic potential as follows
\begin{equation}\label{eq:C6}
    U(\bm{r})=2U\sum_{j=1}^{3}\cos({\bf G}_j\cdot\bf{r}),
\end{equation} 
where $\bm{G}_j=\frac{4\pi}{\sqrt{3}}L^{-1}(-\sin\frac{2\pi j}{3},\cos\frac{2\pi j}{3})$ are three reciprocal vectors, and $U$ is the potential strength. Then there are two energy scales $v_F/L$ and $U$ in Eq. (\ref{eq:dirac}), and the low-energy physics is determined by a single dimensionless control parameter $UL/v_F$.

As shown in the band structure (Fig.~\ref{fig:dirac}b), the first set of satellite Dirac points on positive side are found at $M$ points, and there are generically six saddle points per moir\'e Brillouin zone (MBZ) between the main Dirac point $\Gamma$ and satellite Dirac points $M$. Remarkably, the Fermi surfaces passing through these saddle points are all perfect circles in a wide range of $UL/v_F$ (Fig.~\ref{fig:dirac}c).%, so we name those as Venn-shape Fermi surfaces. 

When $U=1.36 v_F/L$, three ordinary saddle points and one local extremum merge into a high-order saddle point at the $K$ point where all three Fermi surfaces intersect (Fig.~\ref{fig:dirac}c), and the dispersion around the $K$ point becomes flattened  (Fig.~\ref{fig:dirac}b). In the experiment, the potential strength $U$ and the Fermi velocity $v_F$ are mostly determined by the material, but we can tune this parameter $UL/v_F$ by tuning the twisted angle $\theta$ and the resulting superlattice constant $L=a/\theta$. Thus, we can also define a magic angle $\theta_c = 0.74 U a/v_F$ when our system hits high-order VHS, where $a$ is the atomic lattice constant.

\begin{figure}[hbtp]
    \centering
    \includegraphics[width=0.48\textwidth]{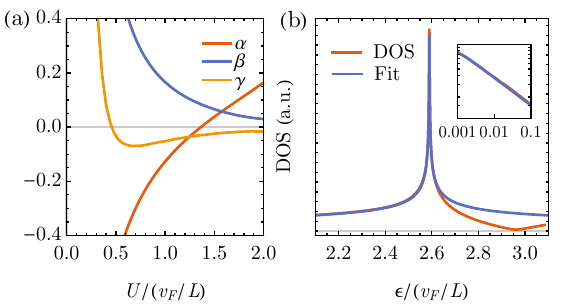}
    \caption{(a) Dimensionless derivatives $\partial^n E/\partial p_{x}^n(v_F L^{n-1}n!)^{-1}$ of the moir\'e surface state dispersion with respect to momentum at the $K$ point, where $\alpha$, $\beta$, and $\gamma$ correspond to $n=2,3,$ and 4 (see Eq.~(\ref{eq:Taylor}) of main text). The potential is given in Eq.~(\ref{eq:C6}). When $UL/v_F=1.36$, $\alpha$ vanishes, making the $K$ point a high-order saddle point. (b) DOS around the high-order VHS. The log-log plot in the inset shows that the divergence of the DOS is indeed power-law. The fitting is given by $N(\epsilon) = c|\epsilon-\epsilon_0|^{\nu}$, where $c = 0.76$, $\epsilon_0 = 2.59$, and $\nu = -0.39$. $\nu = -0.39$ agrees well with theoretical result $\nu = -1/3$ (see main text). }
    \label{fig:highVHS}
\end{figure}

Among high symmetry points $\Gamma,M$ and $K$, the time-reversal invariant points $\Gamma,M$ will always be at least doubly degenerate, while only $K$ point can become spin singlet. We can thus expand the singlet dispersion $E_{\bm{k}}$ around the $K$ point,
\begin{equation}\label{eq:Taylor}
    E_{\bm p+\bm{K}} - E_{\bm{K}} = \alpha p^2 + \beta (p_x^3 - 3p_xp_y^2) +\gamma p^4+ \cdots
\end{equation}
where $p^2=p_x^2 + p_y^2$ with $p_x$ ($p_y$) parallel (perpendicular) to the $\Gamma K$ line. We then compute the Taylor coefficients $\alpha,\beta$ and $\gamma$ as functions of $UL/v_F$ as shown in Fig.~\ref{fig:highVHS}a. When $U = 1.36 v_F/L$, we find $\alpha$ vanishes, while $\beta$ remains finite, indicating a high-order saddle point described by a third-order polynomial $E_{\bm{p}+\bm{K}} - E_{\bm{K}} = \beta (p_x^3 - 3p_xp_y^2)$.

The density of states (DOS) of a $C_3$ saddle point diverges with power-law exponent $\nu = -1/3$ according to the scaling property of the dispersion \cite{highVHS,highCP}.
As shown in Fig.~\ref{fig:highVHS}b, the numerical power-law fitting of DOS gives $\nu = -0.39$, which agrees well with $\nu = -1/3$.

\begin{figure}[htbp]
    \centering
    \includegraphics[width=0.48\textwidth]{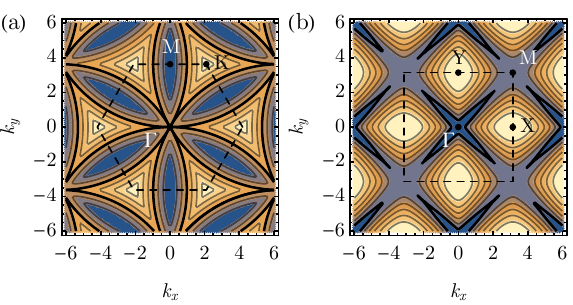}
    \caption{(a) Fermi surface with a high-order Dirac point at the $\Gamma$ point in $C_6$ potential Eq.~(\ref{eq:C6}). (b) Fermi surface with a high-order VHS of class ${A}_2$ on the high-symmetry line $\Gamma M$ in $D_4$ potential Eq.~(\ref{eq:D4}). The ones passing through high-order VHS are plotted in thick black curves, among which those in (a) are perfect circles. The corresponding MBZ is plotted in dashed lines.}
    \label{fig:otherVHS}
\end{figure}

When $U = 0.15 v_F/L$, at energy much higher than the first set of satellite Dirac points, $\Gamma$ point becomes a high-order Dirac point, where six circular Fermi surfaces intersect together as shown in Fig.~\ref{fig:otherVHS}a.
%The six Fermi surfaces are also circular according to numerical evidence.
The high-order Dirac point also exhibits power-law divergent DOS just like high-order VHS.

Next we consider a periodic potential with $D_4$ symmetry
\begin{equation}\label{eq:D4}
    U(\bm{r})=2U[\cos(2\pi x/L)+\cos(2\pi y/L)],
\end{equation}
which corresponds to a square moir\'e superlattice. When $U=4.71 v_F/L$, there are four high-order saddle points on four $\Gamma M$ lines respectively (Fig. \ref{fig:otherVHS}b), where the local dispersion becomes 
$
    E_{\bm{q}+\bm V} - E_{\bm{V}} = aq_{\parallel}^3-bq_{\perp}^2+ \cdots
$ ($ab>0$).
Here $\bm{V}$ denotes the momentum of high-order saddle point, and $q_{\parallel}$ ($q_{\perp}$) is parallel (perpendicular) to the $\Gamma M$ line. Such kind of saddle point can split into at most two critical points: One ordinary saddle point and one ordinary extremum, and we call it a $A_2$ saddle point \cite{highCP}, where the energy contour is beak like (Fig. \ref{fig:otherVHS}b). Details of high-order VHS $A_2$ and high-order Dirac points can be found in Appendix~\ref{sec:otherVHS}. The numerical DOS properties of high-order VHSs agree well with theoretical predictions in Ref.\cite{highVHS}.

In the $D_4$ potential given in Eq.~(\ref{eq:D4}), $U(x+L/2,y+L/2)=-U(x,y)$, hence the system has an additional particle-hole symmetry $E_{\bm k}\to-E_{\bm k}$, which does not exist for $C_6$ potential given in Eq.~(\ref{eq:C6}). This is consistent with the experiment result in bulk \ce{Bi2Te3} crystal, where particle-hole symmetry is broken \cite{Bi2Te3Moire}. In addition, high-symmetry points in the MBZ under the $D_4$ potential are all time-reversal invariant, thus cannot possess nondegenerate high-order VHS.

Although a high-order saddle point requires tuning one parameter, e.g. twist angle, the power-law diverging behavior of the DOS remains present in a wide energy range even when the band structure is not exactly at high-order VHS (Fig.~\ref{fig:UTc}a). Taking our system of moir\'e surface states in $C_6$ potential as an example, the high-order VHS splits into three ordinary VHS and a local max/min under perturbation. Nevertheless, at temperature higher than the energy difference between the VHS and the local max/min, thermal broadened DOS is indistinguishable from the case of high-order VHS. As we will show in the following section, this energy difference is usually tiny in real materials (Fig.~\ref{fig:UTc}b).

In a more realistic model, we may include other ingredients in this system, such as warping effect of surface states and higher order harmonics of the scalar potential, which are beyond our model Eqs.~(\ref{eq:dirac0}) and (\ref{eq:dirac}). 
In momentum space, these effects will result in local perturbations to the energy dispersion near the high-order VHS. 
Consider the $C_3$ saddle point $E=E_{\bm K}+\alpha p^2 + \beta (p_x^3 - 3p_xp_y^2) +\gamma p^4+ \cdots$ at $\bm K$ point as an example, then the perturbations can be described by Taylor series $\Delta E=\Delta E_{\bm K}+\Delta\alpha p^2 +\Delta\beta (p_x^3-3p_xp_y^2)+\Delta\gamma p^4+\dots$ in momentum space. The high-order VHS of the resulting dispersion $E'\equiv E+\Delta E$ is then determined by $\alpha'\equiv\alpha +\Delta\alpha=0$, and hence the critical value of tuning parameter $UL/v_F$ will be perturbed from that in Fig. \ref{fig:highVHS}a.

Such topological argument also applies to other types of high-order VHSs \cite{highCP} in general superlattices. The topology of Brillouin zone (i.e. torus) guarantees the existence of VHS in general lattices \cite{highCP,VHS}. In a moiré superlattice, moreover, the band structure and hence VHS can be manipulated and tuned continuously via mechanical, electrical and other means, such as tuning the twist angle or the gating voltage. Hence, we expect the Hessian matrix determinant of VHS can be tuned to zero with appropriate parameters. As a result, during the continuous tuning of moiré superlattices, we expect ordinary VHS could generally evolve into high-order VHS.

\section{Superconductivity near high-order VHS} \label{sec:SC}

In this section, we consider the physical consequences of high-order VHS. When the chemical potential is put at the energy of high-order VHS, interaction induced instabilities will be greatly enhanced due to the power-law divergent DOS \cite{Chubukov,Chubukov2,Supermetal,Guinea}. In moir\'e surface states of topological insulators, possible interaction channels at high-order VHS are limited due to spin-momentum locking of surface states (i.e., there only exists a single spin-polarized state at every $k$ point on the Fermi surface). In this paper, we focus on attractive interactions in the week coupling regime that is experimentally relevant.

To be specific, at the high order VHS in $C_6$ potential, the divergent density of states occurs near two points $K$ and $K'$ on the Fermi surface, where electron spins are polarized in opposite directions. We denote operators of such states as $c_{K\uparrow}$ and $c_{K'\downarrow}$ respectively, and then the only relevant (in the sense of renormalization group) electron-electron interaction reads $H_{\rm int}=g c_{K\uparrow}^{\dagger} c_{K\uparrow} c_{K'\downarrow}^{\dagger} c_{K'\downarrow}$. Therefore, in this low energy theory, the possible instability due to the attractive interaction $g<0$ is superconductivity. Note that the charge density wave instability that usually coexists with superconductivity is absent due to the opposite spins at $K$ and $K'$, while the spin density wave instability $\langle c_{K\uparrow}^{\dagger} c_{K'\downarrow} \rangle$ is unfavorable under attractive interaction. The absence of moire band gaps and the large Fermi velocity away from VHS also facilitate the reduction of Coulomb repulsion through retardation effects, and thus favor electron-phonon superconductivity.

To find an analytic formula for the superconducting critical temperature $T_c$, we employ the Anderson-Morel approximation to solve the gap equation. We assume the dimensionless interaction takes a simple form with piece wise constant attractive phonon interaction $\lambda>0$ and repulsive interaction $\mu>0$:
\begin{gather} \label{eq:int}
    g(\xi,\xi') = - \lambda \Theta (\xi,\xi')+ \mu,\\
    \Theta (\xi,\xi') = 
    \begin{cases}
    1, & \left|\xi\right|,\left|\xi'\right| < \epsilon_{D}\\
    0, & {\rm otherwise}
    \end{cases},
\end{gather}
where $\xi=E-E_F$ is the electron energy measured from the chemical potential $E_F$.
Notice that $g<0$ means attraction and $g>0$ repulsion.
When we set the chemical potential $E_F$ to be exactly at the high-order VHS, the normalized DOS can be described by the piece wise function
\begin{equation}\label{eq:dos}
    n(\xi) = \begin{cases}
        \left | \Lambda \right |^{-\nu} \left | \xi \right |^{\nu}, \quad &\left|\xi\right| < \Lambda\\
        1, \quad &\Lambda < \left|\xi\right| < W
    \end{cases},
\end{equation}
where $-1 < \nu < 0$ is the power-law exponent of the DOS.
Four energy scales are involved in this problem: the superconducting critical temperature $T_c$, the high-order VHS peak cutoff $\Lambda$, the Debye frequency $\epsilon_D$, and the bandwidth $W$, which satisfy $T_c \ll \Lambda < \epsilon_D < W$. 

Before we get into formal calculations, we first consider various limits with attractive interaction $\lambda$ and simple expression of DOS. The critical temperature $T_c$ is determined by the condition $\lambda\rchi =1$ with the pair susceptibility $\rchi=\int_{T}^{\epsilon_D} n(\xi)\xi^{-1}d\xi$.
When DOS is constant $n(\xi)=1$, pair susceptibility is logarithmically divergent in temperature $\rchi\sim\log(\epsilon_D/T)$, which leads to the BCS formula $T_c\sim\epsilon_D\exp(-1/\lambda)$. When DOS has an ordinary VHS with cutoff $\Lambda$, $n(\xi)=\log(\Lambda/|\xi|)$, we have $\rchi\sim\log^2(\Lambda/T)$ and hence Labb\'e-Bok formula $T_c\sim\Lambda\exp(-1/\sqrt{\lambda})$ \cite{VHSSC,VHSSC2}. When DOS has a high-order VHS with cutoff $\Lambda$ and power-law exponent $\nu$, $\rchi\sim (T/\Lambda)^{\nu}$ and hence we have the power-law formula $T_c\sim\Lambda(1/{\lambda})^{1/\nu}$, where $-1<\nu<0$.

With interaction Eq.~(\ref{eq:int}) and DOS Eq.~(\ref{eq:dos}), we find an exact analytic formula of the critical temperature $T_c$ which generalizes the well known BCS formula,
\begin{equation} \label{eq:tc}
    T_{c} = \frac{\Lambda}{I(\nu)^{1/\nu}} \left [ \frac{1}{\lambda-\mu^*} -  \log (\frac{\epsilon_D}{\Lambda}) + \frac{1}{|\nu|} \right ]^{-1/|\nu|},
\end{equation}
where $\mu^*={\mu}/[{1+\mu \ln \left(W / \epsilon_{D}\right)}]$ is the screened repulsion, $I(\nu) = 2(2^{1-\nu}-1) \Gamma(\nu) \zeta(\nu)$, $\Gamma(\nu)$ and $\zeta(\nu)$ are the gamma function and the zeta function respectively, and we put $\nu = - |\nu|$ to remind the reader $-1<\nu<0$. We direct readers to Appendix~\ref{sec:AndersonMorel} for detailed derivation. 

In the limit $\nu \to 0^-$, the power-law dependence of the effective interaction strength $\lambda^* = \lambda-\mu^*$ disappears because the $1/\nu$ term in the bracket dominates, then Eq.~(\ref{eq:tc}) reduces to the BCS formula of exponential dependence, $ \lim_{\nu\to 0} T_{c} =1.13 \epsilon_D\exp\left (-1/\lambda^*\right )$ (see Appendix~\ref{sec:AndersonMorel}). Reproducing the correct numerical prefactor indicates that our new formula is exact. 

The new formula Eq.~(\ref{eq:tc}) is surprisingly an analytic function of the retarded attractive interaction strength $\lambda^* = \lambda-\mu^*$ with power-law dependence. The analytic nature of the formula suggests a dramatic enhancement of superconductivity compared to ordinary metals and at ordinary VHS when $\lambda^*$ is small. We compare our new formula with previous studies on various DOS in Appendix~\ref{sec:otherTc}.

\begin{figure}[htbp]
    \centering
    \includegraphics[width=0.46\textwidth]{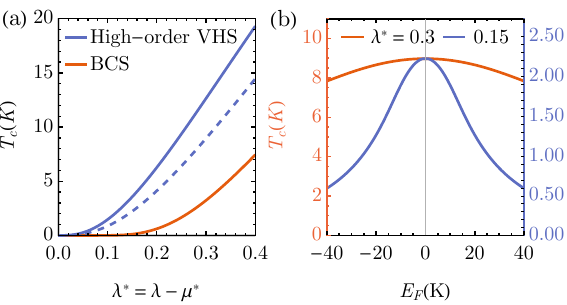}
    \caption{(a) Transition temperature $T_{c}$ at different effective interaction $\lambda^*$ with $\epsilon_D = \SI{80}{\K}$ and $\nu = -1/3$ (the chemical potential is at the high-order VHS). The Anderson-Morel plot refers to the original Anderson-Morel formula $T_c = 1.13 \epsilon_D \exp (- 1/\lambda^*)$ \cite{AndersonMorel}, and the high-order VHS plot refers to Eq.~(\ref{eq:tc}). The solid blue line corresponds to $\Lambda = \epsilon_D$ and the dashed one corresponds to $\Lambda = \epsilon_D/2$, in either case $T_{c}$ with a high-order VHS is much higher than the one without, especially when the effective interaction $\lambda^*$ is small. (b) Transition temperature $T_c$ when the chemical potential $E_F$ is not exactly at the high-order VHS energy with $\Lambda = \epsilon_D/2$. The red axis on the left and the blue axis on the right correspond to the red and blue curve respectively, both of which are $T_c$ in Kelvin. $T_c$ starts to drop when $E_F$ is comparable to $T_c$.}
    \label{fig:Tc}
\end{figure}

We plot the transition temperature $T_{c}$ as a function of effective interaction $\lambda^*$ in Fig.~\ref{fig:Tc}a with parameters relevant to topological insulators. We find that the transition temperature $T_c$ is enhanced enormously by the high-order VHS compared to the original Anderson-Morel result due to the power-law nature of the expression. This enhancement is robust as long as $\Lambda$ and $\epsilon_D$ are at the same order. 
%When $\lambda^* = 0.2$ and $\Lambda = \epsilon_D/2$, the enhancement can still achieve $\gtrapprox 200$ times.

Numerical evidence also shows that the high-order VHS has a robust enhancement effect on superconductivity even when the chemical potential is not exactly at the VHS energy $E_F \neq 0$ (Fig.~\ref{fig:Tc}b). We find that $T_c$ starts to drop when $E_F$ is comparable to $T_c$ and stays more than half of the original $T_c$ even when $E_F$ is five times of $T_c$. This plot can be compared to future experimental data of $T_c$ when varying the filling $E_F$.

\begin{figure}[htbp]
    \centering
    \includegraphics[width=0.48\textwidth]{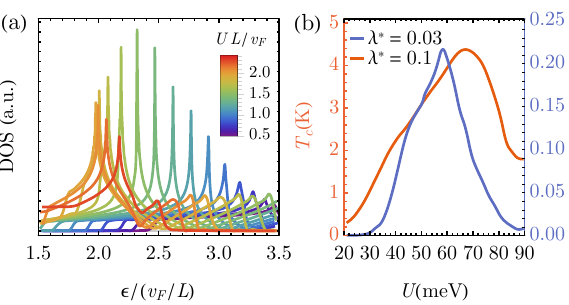}
    \caption{(a) DOS of moir\'e surface states under the $C_6$ potential in Eq.~(\ref{eq:C6}) with various $U$, which has a sharp high-order-VHS-like peak in a wide range of $U$. (b) Transition temperature $T_c$ at different $U$ (the chemical potential is held exactly at the VHS) with $\epsilon_D = \SI{80}{K}$ and $v_F/L = \SI{435}{K}$. The red axis on the left and the blue axis on the right correspond to the red and blue curve respectively, both of which are $T_c$ in Kelvin. $T_c$ shows a wide peak around the potential that corresponds to high-order VHS.}
    \label{fig:UTc}
\end{figure}

Now we come back to the moir\'e surface states in $C_6$ potential to discuss the effect of perturbing the band structure away from high-order VHS. In fact, the DOS has a sharp high-order-VHS-like peak in a wide range of $U$ (Fig.~\ref{fig:UTc}a), which suggests a significant enhancement of superconductivity even when the band structure is not exactly at high-order VHS. We solve the gap equation numerically with the chemical potential held exactly at the VHS. As shown in Fig.~\ref{fig:UTc}b, $T_c$ is enhanced in a wide range of $U$, and the window of $U$ within which $T_c$ is enhanced can be especially wide when $\lambda^*$ is large.
%For example, it is as wide as $\SI{40}{\meV}$ when $\lambda^* > 0.1$.
This also means the energy difference between the VHS and the local max/min is at the order of sub-Kelvin even when the potential is perturbed tens of \SI{}{\meV}s away from the high-order VHS.

Finally we discuss several possible systems to realize our model of Dirac fermion in a periodic potential and the phonon-induced superconductivity within. A prototypical system is the moir\'e topological insulator surface states, where the effective potential can be comparable to the kinetic energy at scale $v_F/L$. In real TIs, the Dirac velocity $v_F$ can be a few \SI{}{\eV \angstrom}'s \cite{Bi2Te3velocity,Bi2Se3velocity,BiSbvelocity}, and the moir\'e supercell constant $L$ can be several or even tens of \SI{}{\nm}'s depending on the lattice mismatch \cite{Bi2Te3Moire,Bi2Se3Moire,Bi2Se3/MoS2DFT,Bi2Se3/MoS2,TMD/Bi2Se3}, then the energy scale $v_F/L$ is at order of tens of \SI{}{\meV}s, which is comparable to the effective potential at moir\'e scale \cite{TBG,TMD,quantumchem}. Furthermore, \ce{Bi2Se3} is believed to have strong electron-phonon interaction \cite{Bi2Se31,Bi2Se32}. Great effort has been put into extract the electron-phonon coupling strength $\lambda$ of \ce{Bi2Se3} both theoretically and experimentally, and most studies fall into the range from $\lambda = 0.2$ to $0.5$ (Fig.~\ref{fig:Tc}) \cite{Sarma,HeliumScattering,ARPES1,ARPES2}. \ce{Bi2Se3} also has a relatively large dielectric constant $\epsilon > 50$ \cite{Bi2Se31,Bi2Se32}, and the bandwidth of topological surface bands reaches at least \SI{800}{meV} \cite{SurfaceStateBandwidth}, so the renormalized Coulomb repulsion $\mu^*$ is negligible, i.e. $\lambda^* \approx \lambda$. With the numbers given, we anticipate that superconductivity can occur on moire surface states of topological insulators with transition temperature up to $T_c \sim \SI{10}{\K}$.

\section{Conclusion}

In this work, we study the moir\'e topological insulator surface states using a continuum model of Dirac electrons moving in periodic potentials at moir\'e scale. Within the continuum model, we identify various types of high-order VHS. We further compute the superconducting transition temperature $T_c$ when the chemical potential is close to the high-order VHS. When exactly at the high-order VHS, we give an analytic formula of $T_c$, showing a power-law instead of exponential dependence of the retarded electron-phonon interaction strength. This result suggests a significantly enhanced superconductivity at high-order VHS, especially when the electron-phonon interaction is weak. In the end, we discuss several real materials that can demonstrate the enhancement of superconductivity due to high-order VHS.

\textit{Note added.--} After this work is completed, we became aware of a related and independent work \cite{Jennifer}.

\begin{acknowledgments}

We acknowledge Hiroki Isobe, Yang Zhang and Yi-Zhuang You for helpful discussion. This work is supported by DOE Office of Basic Energy Sciences, Division of Materials Sciences and Engineering under Award DE-SC0018945. LF is partly supported by the Simons Investigator award from the Simons
Foundation.

\end{acknowledgments}

\begin{appendix}

\section{High-order Dirac point and high-order VHS of type $A_2$} \label{sec:otherVHS}

\begin{figure}[htbp]
    \centering
    \includegraphics[width=0.48\textwidth]{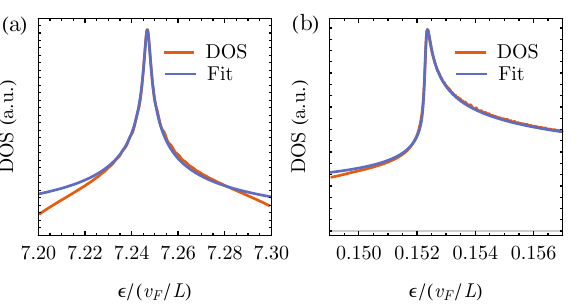}
    \caption{(a) DOS of the relevant bands around the high-order Dirac point. The fitting is given by $N(\epsilon) = c|\epsilon-\epsilon_0|^{\nu} + n$, where $c = 0.047$, $\epsilon_0 = 7.25$, $\nu = -0.33$, and $n = 0.89$. (b) DOS of the relevant band around the high-order VHS of class $A_2$. The fitting is given by $N(\epsilon) = C_{\pm} |\epsilon-\epsilon_0|^{\nu}$ (see Eq.~(\ref{eq:cpm}) for definition) with $C_- = 0.35$, $C_+ = 1.82 C_-$, $\epsilon_0 = 0.152$, and $\nu = -0.20$. Both $\nu$ and $C_+/C_-$ agrees well with the theoretical prediction.}
    \label{fig:OtherDOS}
\end{figure}

\begin{figure*}[htbp]
    \centering
    \includegraphics[width=0.8\textwidth]{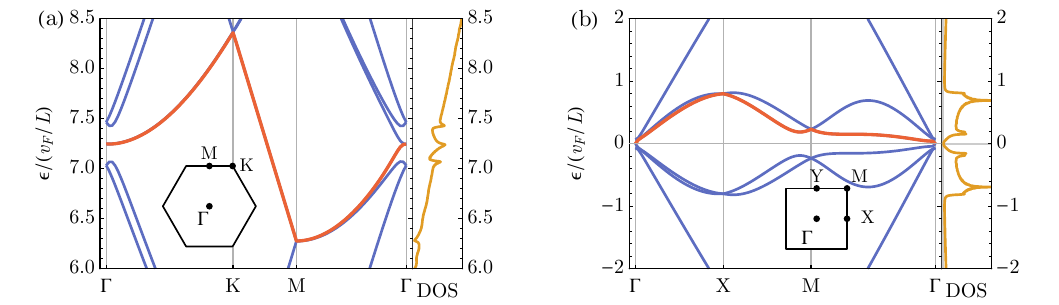}
    \caption{(a) Spectrum and DOS of Dirac fermion in a $C_6$ potential with $U = 0.15 v_F/L$. The high-order Dirac point has been identified at the $\Gamma$ point at energy much higher than the first set of satellite Dirac points (highlighted in red). (b) Spectrum and DOS of Dirac fermion in a $D_4$ potential with $U = 4.71 v_F/L$. The high-order VHS of class $A_2$ has been identified on the high-symmetry line $\Gamma M$ in the primary Dirac cone (highlighted in red).}
    \label{fig:OtherDirac}
\end{figure*}

In this appendix, we describe the high-order Dirac point and high-order VHS of class $A_2$ introduced in the main text in more details. Under $C_6$ potential, the high-order Dirac point is identified at the $\Gamma$ point at energy much higher than the first set of satellite Dirac points. Three secondary Dirac cones in this system are lifted by the $C_6$ potential, and the middle one is in general flattened (Fig.~\ref{fig:OtherDirac}a). When ${U}= 0.15 v_F/L$, the $\Gamma$ point becomes a high-order Dirac point, which also exhibits power-law divergent DOS with $\nu = -1/3$ (Fig.~\ref{fig:OtherDOS}a).

Under $D_4$ potential, the high-order VHS of class $A_2$ is identified on the high-symmetry line $\Gamma M$ in the primary Dirac cone at ${U} = 4.71 v_F/L$ (Fig.~\ref{fig:OtherDirac}b). In this case, the divergence is in perfect agree with the theoretical prediction $\nu = -1/6$ and the asymmetry ratio between the high energy and low energy side is close to $\sqrt{3}$ as predicted in Ref.~\cite{highVHS}. (Fig.~\ref{fig:OtherDOS}b).

\section{Solution to the BCS gap equation at high-order VHS} \label{sec:AndersonMorel}

In this appendix, we solve the BCS gap equation at high-order VHS within Anderson-Morel approximation
\cite{BCS1,BCS2},
\begin{equation}
    \Delta_{k}=-\sum_{k^{\prime}} V_{k, k^{\prime}} \frac{\Delta_{k^{\prime}}}{2 E_{k^{\prime}}} \tanh \frac{E_{k^{\prime}}}{2 T}
\end{equation}
where $\Delta_{k}$ is the superconducting gap, and $E_{k}=\sqrt{\xi_{k}^{2}+\Delta_{k}^{2}}$ is the quasiparticle energy. Due to the complexity of the gap equation, now we proceed within the Anderson-Morel approximation \cite{AndersonMorel}, in which both the interaction strength and the gap are assumed to be piece wise constant. We consider four energy scales: the superconducting gap $\Delta \sim T_c$, the high-order VHS peak cutoff $\Lambda$, the Debye frequency $\epsilon_D$, and the bandwidth $W$, with $T_c \ll \Lambda < \epsilon_D < W$. Now we assume that the dimensionless interaction takes a simple form: piece wise constant attractive phonon interaction $\lambda$ and a repulsive Coloumb interaction $\mu$,
\begin{gather}
    g(\xi,\xi') = - \lambda \Theta (\xi,\xi')+ \mu,\\
    \Theta (\xi,\xi') = 
    \begin{cases}
    1, & \left|\xi\right|,\left|\xi'\right| < \epsilon_{D}\\
    0, & {\rm otherwise}
    \end{cases},
\end{gather}
which is normalized with the constant DOS $N_0$ away from the van Hove singularity, $V_{\xi,\xi'} = g(\xi,\xi')/N_0$. The normalized density of state $n(\xi) = N(\xi)/N_0$ , however, is modified from the original Anderson-Morel model to account for the high-order VHS. For now we set the chemical potential $E_F$ to be exactly at the high-order van Hove singularity,
\begin{equation}
    n(\xi) = \begin{cases}
        C \left | \xi \right |^{\nu}, \quad &\left|\xi\right| < \Lambda\\
        1, \quad &\Lambda < \left|\xi\right| < W
    \end{cases}
\end{equation}
where $-1 < \nu < 0$ is the power-law exponent of the DOS. The continuity condition enforced that $C = \left | \Lambda \right |^{-\nu}$. Within the Anderson-Morel approximation, the gap also takes a simple form,
\begin{equation}
    \Delta(\xi) = \begin{cases}
        \Delta_1, \quad &\left|\xi\right| < \omega_D\\
        \Delta_2, \quad &\omega_D < \left|\xi\right| < W
    \end{cases}
\end{equation}
Plugging in $N(\xi')$ and $V\left(\xi, \xi^{\prime}\right)$, the gap equation becomes
\begin{widetext}
\begin{equation}
    \begin{pmatrix} \Delta_{1} \\ \Delta_{2} \end{pmatrix} =
    \begin{pmatrix} (\lambda-\mu) \left \{ I(\nu) \left ( \frac{T}{\Lambda} \right )^{\nu} + \log (\frac{\epsilon_D}{\Lambda}) +  \frac{1}{\nu} \right \} & -\mu \log \left(W / \epsilon_{D}\right) \\ -\mu \left \{ I(\nu) \left ( \frac{T}{\Lambda} \right )^{\nu} + \log (\frac{\epsilon_D}{\Lambda}) +  \frac{1}{\nu} \right \} & -\mu \log \left(W / \epsilon_{D}\right) \end{pmatrix} \begin{pmatrix} \Delta_{1} \\ \Delta_{2} \end{pmatrix}
\end{equation}
where $I(\nu) = 2(2^{1-\nu}-1) \Gamma(\nu) \zeta(\nu)$, $\Gamma(\nu)$ and $\zeta(\nu)$ are the gamma function and the zeta function respectively. Here we use an important integral,
\begin{equation}
    \int_0^{\epsilon_D} \dd \xi n(\xi) \frac{\tanh \left(\beta \xi / 2\right)}{\xi} = \left ( \frac{T}{\Lambda} \right )^{\nu} \left ( I(\nu) - \int_{\beta \Lambda}^{\infty} \dd x x^{\nu-1} \right ) + \int_{\beta\Lambda}^{\beta\epsilon_D} \frac{\dd x}{x}, \quad \quad
    I(\nu) \equiv \int_{0}^{\infty} \dd x x^{\nu-1} \tanh \left(\frac{x}{2}\right),
\end{equation}
\end{widetext}
where we used $T_c \ll \Lambda$. Solving the consistency equation gives the critical temperature
\begin{equation}
    T_{c} = \frac{\Lambda}{I(\nu)^{1/\nu}} \left [ \frac{1}{\lambda-\mu^*} -  \log (\frac{\epsilon_D}{\Lambda}) + \frac{1}{|\nu|} \right ]^{-1/|\nu|}.
\end{equation}

\section{Comparison of the $T_c$ formula with previous works} \label{sec:otherTc}

\begin{figure*}[htbp]
    \centering
    \includegraphics[width=0.85\textwidth]{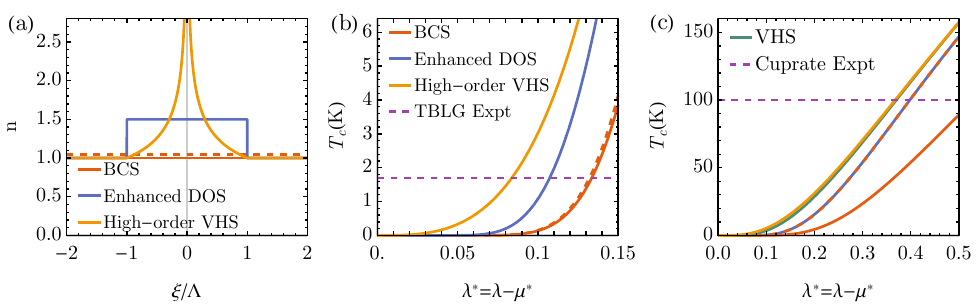}
    \caption{(a) Schematic plot of the DOS in different scenarios ($\nu = -1/3$ for the high-order VHS scenario). The rescaled DOS in the BCS scenario is plotted with a red dashed line with $\epsilon_D = 10 \Lambda$ (see text for definiton). (b-c) Transition temperature $T_c$ at different retarded interaction $\lambda^* = \lambda - \mu^*$ with parameters relevant to (b) twisted bilayer graphene (tBLG) and (c) cuprate superconductor. The red dashed line represents using rescaled DOS in the BCS scenario. Detailed description for each line is in the text. In (b), we pick $\epsilon_D = \SI{230}{\meV}$, $\Lambda = \SI{5}{\meV}$, $\nu = -1/3$. In (c), we pick $\epsilon_D = \SI{50}{\meV}$, $\Lambda = \SI{50}{\meV}$, $\nu = -1/4$.}
    \label{fig:OtherTc}
\end{figure*}

In this appendix, we compare our result with previous works with various DOS (Fig.~\ref{fig:OtherTc}a). In Anderson and Morel's original work, the transition temperature $T_c$ is derived at constant density of states $n(\xi) = 1$ (Fig.~\ref{fig:OtherTc}a),
$
    T_{c} = 1.13 \epsilon_D \exp( -\frac{1}{\lambda-\mu^*} )
$ \cite{AndersonMorel}.
To show how our result reduces to this Anderson-Morel formula in the limit $|\nu| \to 0$, we first use the asymptotic form of the gamma function and the zeta function at $|\nu| \to 0$ to find $I(\nu) \sim - \frac{1}{\nu} + \gamma + \log (2/\pi)$, where $\gamma$ is Euler's constant. Then we plug in to find $T_c$,
\begin{align}
    T_{c} &\sim \Lambda \left ( \frac{n + \gamma + \log (2/\pi)}{n + \frac{1}{\lambda-\mu^*} -  \log (\frac{\epsilon_D}{\Lambda})} \right )^n\\ &\sim \Lambda \left \{ 1 + \log \left  [\frac{2}{\pi} e^{\gamma} \frac{\epsilon_D}{\Lambda} \exp ( - \frac{1}{\lambda-\mu^*} )  \right ] \frac{1}{n} \right \}^n
\end{align}
where $n = 1/|\nu|$. Finally, the definition of Euler's number $e$ simplifies the formula,
\begin{equation}
    \lim_{\nu \to 0^-} T_c = \frac{2}{\pi} e^{\gamma} \epsilon_D \exp ( - \frac{1}{\lambda-\mu^*} ) 
\end{equation}
where $2 e^{\gamma}/ \pi = 1.13$ recovers the correct numerical prefactor in the BCS formula, indicating that our new formula is exact.

Tang \textit{et al.} generalized Anderson and Morel's result to the enhanced DOS scenario in the context of topological crystalline insulator interface superconductivity, where $n(\xi) = \alpha$ within the flat band width $\Lambda$ (Fig.~\ref{fig:OtherTc}a) \cite{AndersonMorelFlat,TangNaturePhysics},
\begin{equation}
    T_{c} = 1.13 \Lambda \left(\frac{\epsilon_D }{\Lambda}\right)^{1/\alpha } \exp ( -\frac{1}{\alpha (\lambda-\mu^*)} )
\end{equation}

The VHS plot refers to Labb\'e and Bok's work in the context of cuprate superconductors in the limit $\Lambda \gg \epsilon_D$, where $n(\xi) = n_1 \log (\Lambda/\xi) + n_0$ within the VHS peak cutoff $\Lambda$ \cite{VHSSC,VHSSC2},
\begin{eqnarray}
    T_{c} = \frac{\Lambda}{2} \exp ( 0.819 + \frac{1}{n_1} - \sqrt{F} ),
\end{eqnarray}
where
$
    F \equiv \left ( \frac{1}{n_1} + 0.819 \right )^2 + \left ( \log (\frac{\epsilon_D}{\Lambda}) \right )^2 - 2 
    -  \frac{2}{n_1} \left ( \log (\frac{2.28 \epsilon_D}{\Lambda}) - \frac{1}{\lambda - \mu^*}\right ).
$
Here $\mu^*$ is renormalized slightly differently from the usual Anderson-Morel screening. We refer the reader to Ref.\cite{VHSSC2} for a more detailed discussion. 

Heikkil{\"a} \textit{et al.} also considered a scenario with power-law divergent DOS $N(\xi) = \xi^{\nu}/\Lambda^{\nu}$ in the context of the multiple Dirac point \cite{Grisha,Tero,Tero2}. Unlike us, they work in the limit $\Lambda \gg \epsilon_D$. In our language, their formula can be rewritten as
\begin{equation}
    T_c = \frac{\left (2 (\nu + 1) \right )^{1/\nu}}{2 J(\nu)^{1/\nu}} \Lambda \left ( \frac{\epsilon_D}{\Lambda} \right )^{1/|\nu| - 1} \left (\frac{1}{\lambda} \right )^{-1/|\nu|},
\end{equation}
where $J(\nu) = \Gamma(-\nu/2)\Gamma((3+\nu)/2)/\sqrt{\pi}$.

Now we compare the transition temperature $T_c$ among these various DOS setups. It is important to normalize the number of states to make a fair comparison. If we normalize the number of states within the bandwidth $W$, the difference among various DOS setup are negligible since $W \gg \Lambda$. If we instead normalize the number of states within the bandwidth $\Lambda$, we find $\alpha = 1/(\nu + 1)$ in the enhanced DOS scenario and $n_1 = -1 + 1/(\nu + 1)$ in the VHS scenario. We also include a rescaled DOS for the BCS scenario by normalizing the number of states within Debye frequency $\epsilon_D$ (Fig.~\ref{fig:OtherTc}a). We plot $T_c$ as a function of the retarded interaction $\lambda^* = \lambda - \mu^*$ in Fig.~\ref{fig:OtherTc}b-c, with parameters relevant for tBLG in Fig.~\ref{fig:OtherTc}b, and those for cuprates in Fig.~\ref{fig:OtherTc}c. In particular, we pick $\nu = -1/3$ for tBLG and $\nu = -1/4$ for cuprates, which corresponds to the leading order high-order VHS for systems with $C_3$ symmetry and $D_4$ symmetry respectively. The tBLG plot is not compatible with the VHS result since Labb\'e and Bok assumes that the flat band range $\Lambda$ is much larger compared to the Debye frequency $\epsilon_D$, which does not hold in tBLG. 

\end{appendix}

\bibliography{main.bib}

%merlin.mbs apsrev4-1.bst 2010-07-25 4.21a (PWD, AO, DPC) hacked
%Control: key (0)
%Control: author (0) dotless jnrlst
%Control: editor formatted (1) identically to author
%Control: production of article title (0) allowed
%Control: page (1) range
%Control: year (0) verbatim
%Control: production of eprint (0) enabled
\begin{thebibliography}{45}%
\makeatletter
\providecommand \@ifxundefined [1]{%
 \@ifx{#1\undefined}
}%
\providecommand \@ifnum [1]{%
 \ifnum #1\expandafter \@firstoftwo
 \else \expandafter \@secondoftwo
 \fi
}%
\providecommand \@ifx [1]{%
 \ifx #1\expandafter \@firstoftwo
 \else \expandafter \@secondoftwo
 \fi
}%
\providecommand \natexlab [1]{#1}%
\providecommand \enquote  [1]{``#1''}%
\providecommand \bibnamefont  [1]{#1}%
\providecommand \bibfnamefont [1]{#1}%
\providecommand \citenamefont [1]{#1}%
\providecommand \href@noop [0]{\@secondoftwo}%
\providecommand \href [0]{\begingroup \@sanitize@url \@href}%
\providecommand \@href[1]{\@@startlink{#1}\@@href}%
\providecommand \@@href[1]{\endgroup#1\@@endlink}%
\providecommand \@sanitize@url [0]{\catcode `\\12\catcode `\$12\catcode
  `\&12\catcode `\#12\catcode `\^12\catcode `\_12\catcode `\%12\relax}%
\providecommand \@@startlink[1]{}%
\providecommand \@@endlink[0]{}%
\providecommand \url  [0]{\begingroup\@sanitize@url \@url }%
\providecommand \@url [1]{\endgroup\@href {#1}{\urlprefix }}%
\providecommand \urlprefix  [0]{URL }%
\providecommand \Eprint [0]{\href }%
\providecommand \doibase [0]{http://dx.doi.org/}%
\providecommand \selectlanguage [0]{\@gobble}%
\providecommand \bibinfo  [0]{\@secondoftwo}%
\providecommand \bibfield  [0]{\@secondoftwo}%
\providecommand \translation [1]{[#1]}%
\providecommand \BibitemOpen [0]{}%
\providecommand \bibitemStop [0]{}%
\providecommand \bibitemNoStop [0]{.\EOS\space}%
\providecommand \EOS [0]{\spacefactor3000\relax}%
\providecommand \BibitemShut  [1]{\csname bibitem#1\endcsname}%
\let\auto@bib@innerbib\@empty
%</preamble>
\bibitem [{\citenamefont {Andrei}\ and\ \citenamefont
  {MacDonald}(2020)}]{MacDonaldReview}%
  \BibitemOpen
  \bibfield  {author} {\bibinfo {author} {\bibfnamefont {Eva~Y.}\ \bibnamefont
  {Andrei}}\ and\ \bibinfo {author} {\bibfnamefont {Allan~H.}\ \bibnamefont
  {MacDonald}},\ }\href@noop {} {\enquote {\bibinfo {title} {Graphene bilayers
  with a twist},}\ } (\bibinfo {year} {2020}),\ \Eprint
  {http://arxiv.org/abs/arXiv:2008.08129} {arXiv:2008.08129} \BibitemShut
  {NoStop}%
\bibitem [{\citenamefont {Balents}\ \emph {et~al.}(2020)\citenamefont
  {Balents}, \citenamefont {Dean}, \citenamefont {Efetov},\ and\ \citenamefont
  {Young}}]{AndreaReview}%
  \BibitemOpen
  \bibfield  {author} {\bibinfo {author} {\bibfnamefont {Leon}\ \bibnamefont
  {Balents}}, \bibinfo {author} {\bibfnamefont {Cory~R.}\ \bibnamefont {Dean}},
  \bibinfo {author} {\bibfnamefont {Dmitri~K.}\ \bibnamefont {Efetov}}, \ and\
  \bibinfo {author} {\bibfnamefont {Andrea~F.}\ \bibnamefont {Young}},\
  }\bibfield  {title} {\enquote {\bibinfo {title} {Superconductivity and strong
  correlations in moir{\'e}flat bands},}\ }\href {\doibase
  10.1038/s41567-020-0906-9} {\bibfield  {journal} {\bibinfo  {journal} {Nature
  Physics}\ }\textbf {\bibinfo {volume} {16}},\ \bibinfo {pages} {725--733}
  (\bibinfo {year} {2020})}\BibitemShut {NoStop}%
\bibitem [{\citenamefont {Carr}\ \emph {et~al.}(2020)\citenamefont {Carr},
  \citenamefont {Fang},\ and\ \citenamefont {Kaxiras}}]{EfthimiosReview}%
  \BibitemOpen
  \bibfield  {author} {\bibinfo {author} {\bibfnamefont {Stephen}\ \bibnamefont
  {Carr}}, \bibinfo {author} {\bibfnamefont {Shiang}\ \bibnamefont {Fang}}, \
  and\ \bibinfo {author} {\bibfnamefont {Efthimios}\ \bibnamefont {Kaxiras}},\
  }\bibfield  {title} {\enquote {\bibinfo {title} {Electronic-structure methods
  for twisted moir{\'e}layers},}\ }\href {\doibase 10.1038/s41578-020-0214-0}
  {\bibfield  {journal} {\bibinfo  {journal} {Nature Reviews Materials}\
  }\textbf {\bibinfo {volume} {5}},\ \bibinfo {pages} {748--763} (\bibinfo
  {year} {2020})}\BibitemShut {NoStop}%
\bibitem [{\citenamefont {Tang}\ and\ \citenamefont
  {Fu}(2014)}]{TangNaturePhysics}%
  \BibitemOpen
  \bibfield  {author} {\bibinfo {author} {\bibfnamefont {Evelyn}\ \bibnamefont
  {Tang}}\ and\ \bibinfo {author} {\bibfnamefont {Liang}\ \bibnamefont {Fu}},\
  }\bibfield  {title} {\enquote {\bibinfo {title} {Strain-induced partially
  flat band, helical snake states and interface superconductivity in
  topological crystalline insulators},}\ }\href {\doibase 10.1038/nphys3109}
  {\bibfield  {journal} {\bibinfo  {journal} {Nature Physics}\ }\textbf
  {\bibinfo {volume} {10}},\ \bibinfo {pages} {964--969} (\bibinfo {year}
  {2014})}\BibitemShut {NoStop}%
\bibitem [{\citenamefont {Liu}\ \emph {et~al.}(2014)\citenamefont {Liu},
  \citenamefont {Li}, \citenamefont {Rajput}, \citenamefont {Gilks},
  \citenamefont {Lari}, \citenamefont {Galindo}, \citenamefont {Weinert},
  \citenamefont {Lazarov},\ and\ \citenamefont {Li}}]{Bi2Se3Moire}%
  \BibitemOpen
  \bibfield  {author} {\bibinfo {author} {\bibfnamefont {Y.}~\bibnamefont
  {Liu}}, \bibinfo {author} {\bibfnamefont {Y.~Y.}\ \bibnamefont {Li}},
  \bibinfo {author} {\bibfnamefont {S.}~\bibnamefont {Rajput}}, \bibinfo
  {author} {\bibfnamefont {D.}~\bibnamefont {Gilks}}, \bibinfo {author}
  {\bibfnamefont {L.}~\bibnamefont {Lari}}, \bibinfo {author} {\bibfnamefont
  {P.~L.}\ \bibnamefont {Galindo}}, \bibinfo {author} {\bibfnamefont
  {M.}~\bibnamefont {Weinert}}, \bibinfo {author} {\bibfnamefont {V.~K.}\
  \bibnamefont {Lazarov}}, \ and\ \bibinfo {author} {\bibfnamefont
  {L.}~\bibnamefont {Li}},\ }\bibfield  {title} {\enquote {\bibinfo {title}
  {Tuning dirac states by strain in the topological insulator
  {Bi$_2$Se$_3$}},}\ }\href {\doibase 10.1038/nphys2898} {\bibfield  {journal}
  {\bibinfo  {journal} {Nature Physics}\ }\textbf {\bibinfo {volume} {10}},\
  \bibinfo {pages} {294--299} (\bibinfo {year} {2014})}\BibitemShut {NoStop}%
\bibitem [{\citenamefont {Schouteden}\ \emph {et~al.}(2016)\citenamefont
  {Schouteden}, \citenamefont {Li}, \citenamefont {Chen}, \citenamefont {Song},
  \citenamefont {Partoens}, \citenamefont {Van~Haesendonck},\ and\
  \citenamefont {Park}}]{Bi2Te3Moire}%
  \BibitemOpen
  \bibfield  {author} {\bibinfo {author} {\bibfnamefont {Koen}\ \bibnamefont
  {Schouteden}}, \bibinfo {author} {\bibfnamefont {Zhe}\ \bibnamefont {Li}},
  \bibinfo {author} {\bibfnamefont {Taishi}\ \bibnamefont {Chen}}, \bibinfo
  {author} {\bibfnamefont {Fengqi}\ \bibnamefont {Song}}, \bibinfo {author}
  {\bibfnamefont {Bart}\ \bibnamefont {Partoens}}, \bibinfo {author}
  {\bibfnamefont {Chris}\ \bibnamefont {Van~Haesendonck}}, \ and\ \bibinfo
  {author} {\bibfnamefont {Kyungwha}\ \bibnamefont {Park}},\ }\bibfield
  {title} {\enquote {\bibinfo {title} {Moir{\'e}superlattices at the
  topological insulator bi2te3},}\ }\href {\doibase 10.1038/srep20278}
  {\bibfield  {journal} {\bibinfo  {journal} {Scientific Reports}\ }\textbf
  {\bibinfo {volume} {6}},\ \bibinfo {pages} {20278} (\bibinfo {year}
  {2016})}\BibitemShut {NoStop}%
\bibitem [{\citenamefont {Vargas}\ \emph {et~al.}(2017)\citenamefont {Vargas},
  \citenamefont {Liu}, \citenamefont {Lane}, \citenamefont {Rubin},
  \citenamefont {Bilgin}, \citenamefont {Hennighausen}, \citenamefont
  {DeCapua}, \citenamefont {Bansil},\ and\ \citenamefont
  {Kar}}]{Bi2Se3/MoS2DFT}%
  \BibitemOpen
  \bibfield  {author} {\bibinfo {author} {\bibfnamefont {Anthony}\ \bibnamefont
  {Vargas}}, \bibinfo {author} {\bibfnamefont {Fangze}\ \bibnamefont {Liu}},
  \bibinfo {author} {\bibfnamefont {Christopher}\ \bibnamefont {Lane}},
  \bibinfo {author} {\bibfnamefont {Daniel}\ \bibnamefont {Rubin}}, \bibinfo
  {author} {\bibfnamefont {Ismail}\ \bibnamefont {Bilgin}}, \bibinfo {author}
  {\bibfnamefont {Zachariah}\ \bibnamefont {Hennighausen}}, \bibinfo {author}
  {\bibfnamefont {Matthew}\ \bibnamefont {DeCapua}}, \bibinfo {author}
  {\bibfnamefont {Arun}\ \bibnamefont {Bansil}}, \ and\ \bibinfo {author}
  {\bibfnamefont {Swastik}\ \bibnamefont {Kar}},\ }\bibfield  {title} {\enquote
  {\bibinfo {title} {Tunable and laser-reconfigurable 2d heterocrystals
  obtained by epitaxial stacking of crystallographically incommensurate
  {Bi$_2$Se$_3$} and {MoS$_2$} atomic layers},}\ }\href {\doibase
  10.1126/sciadv.1601741} {\bibfield  {journal} {\bibinfo  {journal} {Science
  Advances}\ }\textbf {\bibinfo {volume} {3}} (\bibinfo {year} {2017}),\
  10.1126/sciadv.1601741}\BibitemShut {NoStop}%
\bibitem [{\citenamefont {Hennighausen}\ \emph
  {et~al.}(2019{\natexlab{a}})\citenamefont {Hennighausen}, \citenamefont
  {Lane}, \citenamefont {Benabbas}, \citenamefont {Mendez}, \citenamefont
  {Eggenberger}, \citenamefont {Champion}, \citenamefont {Robinson},
  \citenamefont {Bansil},\ and\ \citenamefont {Kar}}]{Bi2Se3/MoS2}%
  \BibitemOpen
  \bibfield  {author} {\bibinfo {author} {\bibfnamefont {Zachariah}\
  \bibnamefont {Hennighausen}}, \bibinfo {author} {\bibfnamefont {Christopher}\
  \bibnamefont {Lane}}, \bibinfo {author} {\bibfnamefont {Abdelkrim}\
  \bibnamefont {Benabbas}}, \bibinfo {author} {\bibfnamefont {Kevin}\
  \bibnamefont {Mendez}}, \bibinfo {author} {\bibfnamefont {Monika}\
  \bibnamefont {Eggenberger}}, \bibinfo {author} {\bibfnamefont {Paul~M.}\
  \bibnamefont {Champion}}, \bibinfo {author} {\bibfnamefont {Jeremy~T.}\
  \bibnamefont {Robinson}}, \bibinfo {author} {\bibfnamefont {Arun}\
  \bibnamefont {Bansil}}, \ and\ \bibinfo {author} {\bibfnamefont {Swastik}\
  \bibnamefont {Kar}},\ }\bibfield  {title} {\enquote {\bibinfo {title}
  {Oxygen-induced in situ manipulation of the interlayer coupling and exciton
  recombination in {Bi$_2$Se$_3$/MoS$_2$} 2d heterostructures},}\ }\bibfield
  {booktitle} {\emph {\bibinfo {booktitle} {ACS Applied Materials \&
  Interfaces}},\ }\href {\doibase 10.1021/acsami.9b02929} {\bibfield  {journal}
  {\bibinfo  {journal} {ACS Applied Materials \& Interfaces}\ }\textbf
  {\bibinfo {volume} {11}},\ \bibinfo {pages} {15913--15921} (\bibinfo {year}
  {2019}{\natexlab{a}})}\BibitemShut {NoStop}%
\bibitem [{\citenamefont {Hennighausen}\ \emph
  {et~al.}(2019{\natexlab{b}})\citenamefont {Hennighausen}, \citenamefont
  {Lane}, \citenamefont {Buda}, \citenamefont {Mathur}, \citenamefont
  {Bansil},\ and\ \citenamefont {Kar}}]{TMD/Bi2Se3}%
  \BibitemOpen
  \bibfield  {author} {\bibinfo {author} {\bibfnamefont {Zachariah}\
  \bibnamefont {Hennighausen}}, \bibinfo {author} {\bibfnamefont {Christopher}\
  \bibnamefont {Lane}}, \bibinfo {author} {\bibfnamefont {Ioana~Gianina}\
  \bibnamefont {Buda}}, \bibinfo {author} {\bibfnamefont {Vineet~K.}\
  \bibnamefont {Mathur}}, \bibinfo {author} {\bibfnamefont {Arun}\ \bibnamefont
  {Bansil}}, \ and\ \bibinfo {author} {\bibfnamefont {Swastik}\ \bibnamefont
  {Kar}},\ }\bibfield  {title} {\enquote {\bibinfo {title} {Evidence of a
  purely electronic two-dimensional lattice at the interface of
  {TMD/Bi$_2$Se$_3$} heterostructures},}\ }\href {\doibase 10.1039/C9NR04412D}
  {\bibfield  {journal} {\bibinfo  {journal} {Nanoscale}\ }\textbf {\bibinfo
  {volume} {11}},\ \bibinfo {pages} {15929--15938} (\bibinfo {year}
  {2019}{\natexlab{b}})}\BibitemShut {NoStop}%
\bibitem [{\citenamefont {Song}\ \emph {et~al.}(2010)\citenamefont {Song},
  \citenamefont {Wang}, \citenamefont {Jiang}, \citenamefont {Zhang},
  \citenamefont {Chang}, \citenamefont {Wang}, \citenamefont {He},
  \citenamefont {Chen}, \citenamefont {Jia}, \citenamefont {Wang},
  \citenamefont {Fang}, \citenamefont {Dai}, \citenamefont {Xie}, \citenamefont
  {Qi}, \citenamefont {Zhang}, \citenamefont {Xue},\ and\ \citenamefont
  {Ma}}]{graphene/Bi2Se3}%
  \BibitemOpen
  \bibfield  {author} {\bibinfo {author} {\bibfnamefont {Can-Li}\ \bibnamefont
  {Song}}, \bibinfo {author} {\bibfnamefont {Yi-Lin}\ \bibnamefont {Wang}},
  \bibinfo {author} {\bibfnamefont {Ye-Ping}\ \bibnamefont {Jiang}}, \bibinfo
  {author} {\bibfnamefont {Yi}~\bibnamefont {Zhang}}, \bibinfo {author}
  {\bibfnamefont {Cui-Zu}\ \bibnamefont {Chang}}, \bibinfo {author}
  {\bibfnamefont {Lili}\ \bibnamefont {Wang}}, \bibinfo {author} {\bibfnamefont
  {Ke}~\bibnamefont {He}}, \bibinfo {author} {\bibfnamefont {Xi}~\bibnamefont
  {Chen}}, \bibinfo {author} {\bibfnamefont {Jin-Feng}\ \bibnamefont {Jia}},
  \bibinfo {author} {\bibfnamefont {Yayu}\ \bibnamefont {Wang}}, \bibinfo
  {author} {\bibfnamefont {Zhong}\ \bibnamefont {Fang}}, \bibinfo {author}
  {\bibfnamefont {Xi}~\bibnamefont {Dai}}, \bibinfo {author} {\bibfnamefont
  {Xin-Cheng}\ \bibnamefont {Xie}}, \bibinfo {author} {\bibfnamefont
  {Xiao-Liang}\ \bibnamefont {Qi}}, \bibinfo {author} {\bibfnamefont
  {Shou-Cheng}\ \bibnamefont {Zhang}}, \bibinfo {author} {\bibfnamefont
  {Qi-Kun}\ \bibnamefont {Xue}}, \ and\ \bibinfo {author} {\bibfnamefont
  {Xucun}\ \bibnamefont {Ma}},\ }\bibfield  {title} {\enquote {\bibinfo {title}
  {Topological insulator {Bi$_2$Se$_3$} thin films grown on double-layer
  graphene by molecular beam epitaxy},}\ }\href {\doibase 10.1063/1.3494595}
  {\bibfield  {journal} {\bibinfo  {journal} {Applied Physics Letters}\
  }\textbf {\bibinfo {volume} {97}},\ \bibinfo {pages} {143118} (\bibinfo
  {year} {2010})}\BibitemShut {NoStop}%
\bibitem [{\citenamefont {Wang}\ \emph {et~al.}(2012)\citenamefont {Wang},
  \citenamefont {Jiang}, \citenamefont {Chen}, \citenamefont {Li},
  \citenamefont {Song}, \citenamefont {Wang}, \citenamefont {He}, \citenamefont
  {Chen}, \citenamefont {Ma},\ and\ \citenamefont {Xue}}]{FeSe/Bi2Se3}%
  \BibitemOpen
  \bibfield  {author} {\bibinfo {author} {\bibfnamefont {Yilin}\ \bibnamefont
  {Wang}}, \bibinfo {author} {\bibfnamefont {Yeping}\ \bibnamefont {Jiang}},
  \bibinfo {author} {\bibfnamefont {Mu}~\bibnamefont {Chen}}, \bibinfo {author}
  {\bibfnamefont {Zhi}\ \bibnamefont {Li}}, \bibinfo {author} {\bibfnamefont
  {Canli}\ \bibnamefont {Song}}, \bibinfo {author} {\bibfnamefont {Lili}\
  \bibnamefont {Wang}}, \bibinfo {author} {\bibfnamefont {Ke}~\bibnamefont
  {He}}, \bibinfo {author} {\bibfnamefont {Xi}~\bibnamefont {Chen}}, \bibinfo
  {author} {\bibfnamefont {Xucun}\ \bibnamefont {Ma}}, \ and\ \bibinfo {author}
  {\bibfnamefont {Qi-Kun}\ \bibnamefont {Xue}},\ }\bibfield  {title} {\enquote
  {\bibinfo {title} {Scanning tunneling microscopy of interface properties of
  {Bi$_2$Se$_3$} on {FeSe}},}\ }\href {\doibase 10.1088/0953-8984/24/47/475604}
  {\bibfield  {journal} {\bibinfo  {journal} {Journal of Physics: Condensed
  Matter}\ }\textbf {\bibinfo {volume} {24}},\ \bibinfo {pages} {475604}
  (\bibinfo {year} {2012})}\BibitemShut {NoStop}%
\bibitem [{\citenamefont {Jeon}\ \emph {et~al.}(2011)\citenamefont {Jeon},
  \citenamefont {Jang}, \citenamefont {Yoon},\ and\ \citenamefont
  {Kahng}}]{Au/Bi2Se3}%
  \BibitemOpen
  \bibfield  {author} {\bibinfo {author} {\bibfnamefont {Jeong~Heum}\
  \bibnamefont {Jeon}}, \bibinfo {author} {\bibfnamefont {Won~Jun}\
  \bibnamefont {Jang}}, \bibinfo {author} {\bibfnamefont {Jong~Keon}\
  \bibnamefont {Yoon}}, \ and\ \bibinfo {author} {\bibfnamefont {Se-Jong}\
  \bibnamefont {Kahng}},\ }\bibfield  {title} {\enquote {\bibinfo {title}
  {Metal-supported high crystalline {Bi$_2$Se$_3$} quintuple layers},}\ }\href
  {\doibase 10.1088/0957-4484/22/46/465602} {\bibfield  {journal} {\bibinfo
  {journal} {Nanotechnology}\ }\textbf {\bibinfo {volume} {22}},\ \bibinfo
  {pages} {465602} (\bibinfo {year} {2011})}\BibitemShut {NoStop}%
\bibitem [{\citenamefont {Xu}\ \emph {et~al.}(2015{\natexlab{a}})\citenamefont
  {Xu}, \citenamefont {Han}, \citenamefont {Chen}, \citenamefont {Wu},
  \citenamefont {Wang}, \citenamefont {Han}, \citenamefont {Ye}, \citenamefont
  {Lu}, \citenamefont {Long}, \citenamefont {Wu}, \citenamefont {Lin},
  \citenamefont {Cai}, \citenamefont {Ho}, \citenamefont {He},\ and\
  \citenamefont {Wang}}]{hBN/Bi2Se3}%
  \BibitemOpen
  \bibfield  {author} {\bibinfo {author} {\bibfnamefont {Shuigang}\
  \bibnamefont {Xu}}, \bibinfo {author} {\bibfnamefont {Yu}~\bibnamefont
  {Han}}, \bibinfo {author} {\bibfnamefont {Xiaolong}\ \bibnamefont {Chen}},
  \bibinfo {author} {\bibfnamefont {Zefei}\ \bibnamefont {Wu}}, \bibinfo
  {author} {\bibfnamefont {Lin}\ \bibnamefont {Wang}}, \bibinfo {author}
  {\bibfnamefont {Tianyi}\ \bibnamefont {Han}}, \bibinfo {author}
  {\bibfnamefont {Weiguang}\ \bibnamefont {Ye}}, \bibinfo {author}
  {\bibfnamefont {Huanhuan}\ \bibnamefont {Lu}}, \bibinfo {author}
  {\bibfnamefont {Gen}\ \bibnamefont {Long}}, \bibinfo {author} {\bibfnamefont
  {Yingying}\ \bibnamefont {Wu}}, \bibinfo {author} {\bibfnamefont
  {Jiangxiazi}\ \bibnamefont {Lin}}, \bibinfo {author} {\bibfnamefont {Yuan}\
  \bibnamefont {Cai}}, \bibinfo {author} {\bibfnamefont {K.~M.}\ \bibnamefont
  {Ho}}, \bibinfo {author} {\bibfnamefont {Yuheng}\ \bibnamefont {He}}, \ and\
  \bibinfo {author} {\bibfnamefont {Ning}\ \bibnamefont {Wang}},\ }\bibfield
  {title} {\enquote {\bibinfo {title} {van der {Waals} epitaxial growth of
  atomically thin {Bi$_2$Se$_3$} and thickness-dependent topological phase
  transition},}\ }\bibfield  {booktitle} {\emph {\bibinfo {booktitle} {Nano
  Letters}},\ }\href {\doibase 10.1021/acs.nanolett.5b00247} {\bibfield
  {journal} {\bibinfo  {journal} {Nano Letters}\ }\textbf {\bibinfo {volume}
  {15}},\ \bibinfo {pages} {2645--2651} (\bibinfo {year}
  {2015}{\natexlab{a}})}\BibitemShut {NoStop}%
\bibitem [{\citenamefont {Xu}\ \emph {et~al.}(2015{\natexlab{b}})\citenamefont
  {Xu}, \citenamefont {Wang}, \citenamefont {Liu}, \citenamefont {Ge},
  \citenamefont {Yang}, \citenamefont {Liu}, \citenamefont {Xu}, \citenamefont
  {Guan}, \citenamefont {Gao}, \citenamefont {Qian}, \citenamefont {Liu},
  \citenamefont {Wang}, \citenamefont {Zhang}, \citenamefont {Xue},\ and\
  \citenamefont {Jia}}]{NbSe/Bi2Se3}%
  \BibitemOpen
  \bibfield  {author} {\bibinfo {author} {\bibfnamefont {Jin-Peng}\
  \bibnamefont {Xu}}, \bibinfo {author} {\bibfnamefont {Mei-Xiao}\ \bibnamefont
  {Wang}}, \bibinfo {author} {\bibfnamefont {Zhi~Long}\ \bibnamefont {Liu}},
  \bibinfo {author} {\bibfnamefont {Jian-Feng}\ \bibnamefont {Ge}}, \bibinfo
  {author} {\bibfnamefont {Xiaojun}\ \bibnamefont {Yang}}, \bibinfo {author}
  {\bibfnamefont {Canhua}\ \bibnamefont {Liu}}, \bibinfo {author}
  {\bibfnamefont {Zhu~An}\ \bibnamefont {Xu}}, \bibinfo {author} {\bibfnamefont
  {Dandan}\ \bibnamefont {Guan}}, \bibinfo {author} {\bibfnamefont {Chun~Lei}\
  \bibnamefont {Gao}}, \bibinfo {author} {\bibfnamefont {Dong}\ \bibnamefont
  {Qian}}, \bibinfo {author} {\bibfnamefont {Ying}\ \bibnamefont {Liu}},
  \bibinfo {author} {\bibfnamefont {Qiang-Hua}\ \bibnamefont {Wang}}, \bibinfo
  {author} {\bibfnamefont {Fu-Chun}\ \bibnamefont {Zhang}}, \bibinfo {author}
  {\bibfnamefont {Qi-Kun}\ \bibnamefont {Xue}}, \ and\ \bibinfo {author}
  {\bibfnamefont {Jin-Feng}\ \bibnamefont {Jia}},\ }\bibfield  {title}
  {\enquote {\bibinfo {title} {Experimental detection of a majorana mode in the
  core of a magnetic vortex inside a topological insulator-superconductor
  {Bi$_2$Te$_3$/NbSe$_2$} heterostructure},}\ }\href {\doibase
  10.1103/PhysRevLett.114.017001} {\bibfield  {journal} {\bibinfo  {journal}
  {Phys. Rev. Lett.}\ }\textbf {\bibinfo {volume} {114}},\ \bibinfo {pages}
  {017001} (\bibinfo {year} {2015}{\natexlab{b}})}\BibitemShut {NoStop}%
\bibitem [{\citenamefont {Yuan}\ \emph {et~al.}(2019)\citenamefont {Yuan},
  \citenamefont {Isobe},\ and\ \citenamefont {Fu}}]{highVHS}%
  \BibitemOpen
  \bibfield  {author} {\bibinfo {author} {\bibfnamefont {Noah F.~Q.}\
  \bibnamefont {Yuan}}, \bibinfo {author} {\bibfnamefont {Hiroki}\ \bibnamefont
  {Isobe}}, \ and\ \bibinfo {author} {\bibfnamefont {Liang}\ \bibnamefont
  {Fu}},\ }\bibfield  {title} {\enquote {\bibinfo {title} {Magic of high-order
  van hove singularity},}\ }\href {\doibase 10.1038/s41467-019-13670-9}
  {\bibfield  {journal} {\bibinfo  {journal} {Nature Communications}\ }\textbf
  {\bibinfo {volume} {10}},\ \bibinfo {pages} {5769} (\bibinfo {year}
  {2019})}\BibitemShut {NoStop}%
\bibitem [{\citenamefont {Nandkishore}\ \emph {et~al.}(2012)\citenamefont
  {Nandkishore}, \citenamefont {Levitov},\ and\ \citenamefont
  {Chubukov}}]{Chubukov}%
  \BibitemOpen
  \bibfield  {author} {\bibinfo {author} {\bibfnamefont {Rahul}\ \bibnamefont
  {Nandkishore}}, \bibinfo {author} {\bibfnamefont {L.~S.}\ \bibnamefont
  {Levitov}}, \ and\ \bibinfo {author} {\bibfnamefont {A.~V.}\ \bibnamefont
  {Chubukov}},\ }\bibfield  {title} {\enquote {\bibinfo {title} {Chiral
  superconductivity from repulsive interactions in doped graphene},}\ }\href
  {\doibase 10.1038/nphys2208} {\bibfield  {journal} {\bibinfo  {journal}
  {Nature Physics}\ }\textbf {\bibinfo {volume} {8}},\ \bibinfo {pages}
  {158--163} (\bibinfo {year} {2012})}\BibitemShut {NoStop}%
\bibitem [{\citenamefont {Nandkishore}\ and\ \citenamefont
  {Chubukov}(2012)}]{Chubukov2}%
  \BibitemOpen
  \bibfield  {author} {\bibinfo {author} {\bibfnamefont {Rahul}\ \bibnamefont
  {Nandkishore}}\ and\ \bibinfo {author} {\bibfnamefont {Andrey~V.}\
  \bibnamefont {Chubukov}},\ }\bibfield  {title} {\enquote {\bibinfo {title}
  {Interplay of superconductivity and spin-density-wave order in doped
  graphene},}\ }\href {\doibase 10.1103/PhysRevB.86.115426} {\bibfield
  {journal} {\bibinfo  {journal} {Phys. Rev. B}\ }\textbf {\bibinfo {volume}
  {86}},\ \bibinfo {pages} {115426} (\bibinfo {year} {2012})}\BibitemShut
  {NoStop}%
\bibitem [{\citenamefont {{Isobe}}\ and\ \citenamefont
  {{Fu}}(2019)}]{Supermetal}%
  \BibitemOpen
  \bibfield  {author} {\bibinfo {author} {\bibfnamefont {Hiroki}\ \bibnamefont
  {{Isobe}}}\ and\ \bibinfo {author} {\bibfnamefont {Liang}\ \bibnamefont
  {{Fu}}},\ }\bibfield  {title} {\enquote {\bibinfo {title} {{Supermetal}},}\
  }\href@noop {} {\bibfield  {journal} {\bibinfo  {journal} {arXiv e-prints}\
  ,\ \bibinfo {eid} {arXiv:1905.05188}} (\bibinfo {year} {2019})},\ \Eprint
  {http://arxiv.org/abs/1905.05188} {arXiv:1905.05188 [cond-mat.str-el]}
  \BibitemShut {NoStop}%
\bibitem [{\citenamefont {Guinea}\ and\ \citenamefont {Walet}(2018)}]{Guinea}%
  \BibitemOpen
  \bibfield  {author} {\bibinfo {author} {\bibfnamefont {Francisco}\
  \bibnamefont {Guinea}}\ and\ \bibinfo {author} {\bibfnamefont {Niels~R.}\
  \bibnamefont {Walet}},\ }\bibfield  {title} {\enquote {\bibinfo {title}
  {Electrostatic effects, band distortions, and superconductivity in twisted
  graphene bilayers},}\ }\href {\doibase 10.1073/pnas.1810947115} {\bibfield
  {journal} {\bibinfo  {journal} {Proceedings of the National Academy of
  Sciences}\ }\textbf {\bibinfo {volume} {115}},\ \bibinfo {pages}
  {13174--13179} (\bibinfo {year} {2018})},\ \Eprint
  {http://arxiv.org/abs/https://www.pnas.org/content/115/52/13174.full.pdf}
  {https://www.pnas.org/content/115/52/13174.full.pdf} \BibitemShut {NoStop}%
\bibitem [{\citenamefont {Labb{\'{e}}}\ and\ \citenamefont
  {Bok}(1987)}]{VHSSC}%
  \BibitemOpen
  \bibfield  {author} {\bibinfo {author} {\bibfnamefont {J}~\bibnamefont
  {Labb{\'{e}}}}\ and\ \bibinfo {author} {\bibfnamefont {J}~\bibnamefont
  {Bok}},\ }\bibfield  {title} {\enquote {\bibinfo {title} {Superconductivity
  in alcaline-earth-substituted {La$_2$CuO$_4$}: A theoretical model},}\ }\href
  {\doibase 10.1209/0295-5075/3/11/012} {\bibfield  {journal} {\bibinfo
  {journal} {Europhysics Letters ({EPL})}\ }\textbf {\bibinfo {volume} {3}},\
  \bibinfo {pages} {1225--1230} (\bibinfo {year} {1987})}\BibitemShut {NoStop}%
\bibitem [{\citenamefont {Bok}(1993)}]{VHSSC2}%
  \BibitemOpen
  \bibfield  {author} {\bibinfo {author} {\bibfnamefont {J.}~\bibnamefont
  {Bok}},\ }\bibfield  {title} {\enquote {\bibinfo {title} {Superconductivity
  in the cuprates. the van hove scenario},}\ }\href {\doibase
  https://doi.org/10.1016/0921-4534(93)90882-Q} {\bibfield  {journal} {\bibinfo
   {journal} {Physica C: Superconductivity}\ }\textbf {\bibinfo {volume}
  {209}},\ \bibinfo {pages} {107 -- 112} (\bibinfo {year} {1993})}\BibitemShut
  {NoStop}%
\bibitem [{\citenamefont {Morel}\ and\ \citenamefont
  {Anderson}(1962)}]{AndersonMorel}%
  \BibitemOpen
  \bibfield  {author} {\bibinfo {author} {\bibfnamefont {P.}~\bibnamefont
  {Morel}}\ and\ \bibinfo {author} {\bibfnamefont {P.~W.}\ \bibnamefont
  {Anderson}},\ }\bibfield  {title} {\enquote {\bibinfo {title} {Calculation of
  the superconducting state parameters with retarded electron-phonon
  interaction},}\ }\href {\doibase 10.1103/PhysRev.125.1263} {\bibfield
  {journal} {\bibinfo  {journal} {Phys. Rev.}\ }\textbf {\bibinfo {volume}
  {125}},\ \bibinfo {pages} {1263--1271} (\bibinfo {year} {1962})}\BibitemShut
  {NoStop}%
\bibitem [{\citenamefont {Park}\ \emph {et~al.}(2008)\citenamefont {Park},
  \citenamefont {Yang}, \citenamefont {Son}, \citenamefont {Cohen},\ and\
  \citenamefont {Louie}}]{Graphene}%
  \BibitemOpen
  \bibfield  {author} {\bibinfo {author} {\bibfnamefont {Cheol-Hwan}\
  \bibnamefont {Park}}, \bibinfo {author} {\bibfnamefont {Li}~\bibnamefont
  {Yang}}, \bibinfo {author} {\bibfnamefont {Young-Woo}\ \bibnamefont {Son}},
  \bibinfo {author} {\bibfnamefont {Marvin~L.}\ \bibnamefont {Cohen}}, \ and\
  \bibinfo {author} {\bibfnamefont {Steven~G.}\ \bibnamefont {Louie}},\
  }\bibfield  {title} {\enquote {\bibinfo {title} {Anisotropic behaviours of
  massless dirac fermions in graphene under periodic potentials},}\ }\href
  {\doibase 10.1038/nphys890} {\bibfield  {journal} {\bibinfo  {journal}
  {Nature Physics}\ }\textbf {\bibinfo {volume} {4}},\ \bibinfo {pages}
  {213--217} (\bibinfo {year} {2008})}\BibitemShut {NoStop}%
\bibitem [{\citenamefont {Yuan}\ and\ \citenamefont {Fu}(2020)}]{highCP}%
  \BibitemOpen
  \bibfield  {author} {\bibinfo {author} {\bibfnamefont {Noah F.~Q.}\
  \bibnamefont {Yuan}}\ and\ \bibinfo {author} {\bibfnamefont {Liang}\
  \bibnamefont {Fu}},\ }\bibfield  {title} {\enquote {\bibinfo {title}
  {Classification of critical points in energy bands based on topology,
  scaling, and symmetry},}\ }\href {\doibase 10.1103/PhysRevB.101.125120}
  {\bibfield  {journal} {\bibinfo  {journal} {Phys. Rev. B}\ }\textbf {\bibinfo
  {volume} {101}},\ \bibinfo {pages} {125120} (\bibinfo {year}
  {2020})}\BibitemShut {NoStop}%
\bibitem [{\citenamefont {Van~Hove}(1953)}]{VHS}%
  \BibitemOpen
  \bibfield  {author} {\bibinfo {author} {\bibfnamefont {L\'eon}\ \bibnamefont
  {Van~Hove}},\ }\bibfield  {title} {\enquote {\bibinfo {title} {The occurrence
  of singularities in the elastic frequency distribution of a crystal},}\
  }\href {\doibase 10.1103/PhysRev.89.1189} {\bibfield  {journal} {\bibinfo
  {journal} {Phys. Rev.}\ }\textbf {\bibinfo {volume} {89}},\ \bibinfo {pages}
  {1189--1193} (\bibinfo {year} {1953})}\BibitemShut {NoStop}%
\bibitem [{\citenamefont {Chen}\ \emph {et~al.}(2009)\citenamefont {Chen},
  \citenamefont {Analytis}, \citenamefont {Chu}, \citenamefont {Liu},
  \citenamefont {Mo}, \citenamefont {Qi}, \citenamefont {Zhang}, \citenamefont
  {Lu}, \citenamefont {Dai}, \citenamefont {Fang}, \citenamefont {Zhang},
  \citenamefont {Fisher}, \citenamefont {Hussain},\ and\ \citenamefont
  {Shen}}]{Bi2Te3velocity}%
  \BibitemOpen
  \bibfield  {author} {\bibinfo {author} {\bibfnamefont {Y.~L.}\ \bibnamefont
  {Chen}}, \bibinfo {author} {\bibfnamefont {J.~G.}\ \bibnamefont {Analytis}},
  \bibinfo {author} {\bibfnamefont {J.-H.}\ \bibnamefont {Chu}}, \bibinfo
  {author} {\bibfnamefont {Z.~K.}\ \bibnamefont {Liu}}, \bibinfo {author}
  {\bibfnamefont {S.-K.}\ \bibnamefont {Mo}}, \bibinfo {author} {\bibfnamefont
  {X.~L.}\ \bibnamefont {Qi}}, \bibinfo {author} {\bibfnamefont {H.~J.}\
  \bibnamefont {Zhang}}, \bibinfo {author} {\bibfnamefont {D.~H.}\ \bibnamefont
  {Lu}}, \bibinfo {author} {\bibfnamefont {X.}~\bibnamefont {Dai}}, \bibinfo
  {author} {\bibfnamefont {Z.}~\bibnamefont {Fang}}, \bibinfo {author}
  {\bibfnamefont {S.~C.}\ \bibnamefont {Zhang}}, \bibinfo {author}
  {\bibfnamefont {I.~R.}\ \bibnamefont {Fisher}}, \bibinfo {author}
  {\bibfnamefont {Z.}~\bibnamefont {Hussain}}, \ and\ \bibinfo {author}
  {\bibfnamefont {Z.-X.}\ \bibnamefont {Shen}},\ }\bibfield  {title} {\enquote
  {\bibinfo {title} {Experimental realization of a three-dimensional
  topological insulator, {Bi$_2$Te$_3$}},}\ }\href {\doibase
  10.1126/science.1173034} {\bibfield  {journal} {\bibinfo  {journal}
  {Science}\ }\textbf {\bibinfo {volume} {325}},\ \bibinfo {pages} {178--181}
  (\bibinfo {year} {2009})}\BibitemShut {NoStop}%
\bibitem [{\citenamefont {Xia}\ \emph {et~al.}(2009{\natexlab{a}})\citenamefont
  {Xia}, \citenamefont {Qian}, \citenamefont {Hsieh}, \citenamefont {Wray},
  \citenamefont {Pal}, \citenamefont {Lin}, \citenamefont {Bansil},
  \citenamefont {Grauer}, \citenamefont {Hor}, \citenamefont {Cava},\ and\
  \citenamefont {Hasan}}]{Bi2Se3velocity}%
  \BibitemOpen
  \bibfield  {author} {\bibinfo {author} {\bibfnamefont {Y.}~\bibnamefont
  {Xia}}, \bibinfo {author} {\bibfnamefont {D.}~\bibnamefont {Qian}}, \bibinfo
  {author} {\bibfnamefont {D.}~\bibnamefont {Hsieh}}, \bibinfo {author}
  {\bibfnamefont {L.}~\bibnamefont {Wray}}, \bibinfo {author} {\bibfnamefont
  {A.}~\bibnamefont {Pal}}, \bibinfo {author} {\bibfnamefont {H.}~\bibnamefont
  {Lin}}, \bibinfo {author} {\bibfnamefont {A.}~\bibnamefont {Bansil}},
  \bibinfo {author} {\bibfnamefont {D.}~\bibnamefont {Grauer}}, \bibinfo
  {author} {\bibfnamefont {Y.~S.}\ \bibnamefont {Hor}}, \bibinfo {author}
  {\bibfnamefont {R.~J.}\ \bibnamefont {Cava}}, \ and\ \bibinfo {author}
  {\bibfnamefont {M.~Z.}\ \bibnamefont {Hasan}},\ }\bibfield  {title} {\enquote
  {\bibinfo {title} {Observation of a large-gap topological-insulator class
  with a single dirac cone on the surface},}\ }\href {\doibase
  10.1038/nphys1274} {\bibfield  {journal} {\bibinfo  {journal} {Nature
  Physics}\ }\textbf {\bibinfo {volume} {5}},\ \bibinfo {pages} {398--402}
  (\bibinfo {year} {2009}{\natexlab{a}})}\BibitemShut {NoStop}%
\bibitem [{\citenamefont {Hsieh}\ \emph {et~al.}(2008)\citenamefont {Hsieh},
  \citenamefont {Qian}, \citenamefont {Wray}, \citenamefont {Xia},
  \citenamefont {Hor}, \citenamefont {Cava},\ and\ \citenamefont
  {Hasan}}]{BiSbvelocity}%
  \BibitemOpen
  \bibfield  {author} {\bibinfo {author} {\bibfnamefont {D.}~\bibnamefont
  {Hsieh}}, \bibinfo {author} {\bibfnamefont {D.}~\bibnamefont {Qian}},
  \bibinfo {author} {\bibfnamefont {L.}~\bibnamefont {Wray}}, \bibinfo {author}
  {\bibfnamefont {Y.}~\bibnamefont {Xia}}, \bibinfo {author} {\bibfnamefont
  {Y.~S.}\ \bibnamefont {Hor}}, \bibinfo {author} {\bibfnamefont {R.~J.}\
  \bibnamefont {Cava}}, \ and\ \bibinfo {author} {\bibfnamefont {M.~Z.}\
  \bibnamefont {Hasan}},\ }\bibfield  {title} {\enquote {\bibinfo {title} {A
  topological dirac insulator in a quantum spin hall phase},}\ }\href {\doibase
  10.1038/nature06843} {\bibfield  {journal} {\bibinfo  {journal} {Nature}\
  }\textbf {\bibinfo {volume} {452}},\ \bibinfo {pages} {970--974} (\bibinfo
  {year} {2008})}\BibitemShut {NoStop}%
\bibitem [{\citenamefont {Kerelsky}\ \emph {et~al.}(2019)\citenamefont
  {Kerelsky}, \citenamefont {McGilly}, \citenamefont {Kennes}, \citenamefont
  {Xian}, \citenamefont {Yankowitz}, \citenamefont {Chen}, \citenamefont
  {Watanabe}, \citenamefont {Taniguchi}, \citenamefont {Hone}, \citenamefont
  {Dean}, \citenamefont {Rubio},\ and\ \citenamefont {Pasupathy}}]{TBG}%
  \BibitemOpen
  \bibfield  {author} {\bibinfo {author} {\bibfnamefont {Alexander}\
  \bibnamefont {Kerelsky}}, \bibinfo {author} {\bibfnamefont {Leo~J.}\
  \bibnamefont {McGilly}}, \bibinfo {author} {\bibfnamefont {Dante~M.}\
  \bibnamefont {Kennes}}, \bibinfo {author} {\bibfnamefont {Lede}\ \bibnamefont
  {Xian}}, \bibinfo {author} {\bibfnamefont {Matthew}\ \bibnamefont
  {Yankowitz}}, \bibinfo {author} {\bibfnamefont {Shaowen}\ \bibnamefont
  {Chen}}, \bibinfo {author} {\bibfnamefont {K.}~\bibnamefont {Watanabe}},
  \bibinfo {author} {\bibfnamefont {T.}~\bibnamefont {Taniguchi}}, \bibinfo
  {author} {\bibfnamefont {James}\ \bibnamefont {Hone}}, \bibinfo {author}
  {\bibfnamefont {Cory}\ \bibnamefont {Dean}}, \bibinfo {author} {\bibfnamefont
  {Angel}\ \bibnamefont {Rubio}}, \ and\ \bibinfo {author} {\bibfnamefont
  {Abhay~N.}\ \bibnamefont {Pasupathy}},\ }\bibfield  {title} {\enquote
  {\bibinfo {title} {Maximized electron interactions at the magic angle in
  twisted bilayer graphene},}\ }\href {\doibase 10.1038/s41586-019-1431-9}
  {\bibfield  {journal} {\bibinfo  {journal} {Nature}\ }\textbf {\bibinfo
  {volume} {572}},\ \bibinfo {pages} {95--100} (\bibinfo {year}
  {2019})}\BibitemShut {NoStop}%
\bibitem [{\citenamefont {Wu}\ \emph {et~al.}(2019)\citenamefont {Wu},
  \citenamefont {Lovorn}, \citenamefont {Tutuc}, \citenamefont {Martin},\ and\
  \citenamefont {MacDonald}}]{TMD}%
  \BibitemOpen
  \bibfield  {author} {\bibinfo {author} {\bibfnamefont {Fengcheng}\
  \bibnamefont {Wu}}, \bibinfo {author} {\bibfnamefont {Timothy}\ \bibnamefont
  {Lovorn}}, \bibinfo {author} {\bibfnamefont {Emanuel}\ \bibnamefont {Tutuc}},
  \bibinfo {author} {\bibfnamefont {Ivar}\ \bibnamefont {Martin}}, \ and\
  \bibinfo {author} {\bibfnamefont {A.~H.}\ \bibnamefont {MacDonald}},\
  }\bibfield  {title} {\enquote {\bibinfo {title} {Topological insulators in
  twisted transition metal dichalcogenide homobilayers},}\ }\href {\doibase
  10.1103/PhysRevLett.122.086402} {\bibfield  {journal} {\bibinfo  {journal}
  {Phys. Rev. Lett.}\ }\textbf {\bibinfo {volume} {122}},\ \bibinfo {pages}
  {086402} (\bibinfo {year} {2019})}\BibitemShut {NoStop}%
\bibitem [{\citenamefont {Zhang}\ \emph {et~al.}(2019)\citenamefont {Zhang},
  \citenamefont {Yuan},\ and\ \citenamefont {Fu}}]{quantumchem}%
  \BibitemOpen
  \bibfield  {author} {\bibinfo {author} {\bibfnamefont {Yang}\ \bibnamefont
  {Zhang}}, \bibinfo {author} {\bibfnamefont {Noah F.~Q.}\ \bibnamefont
  {Yuan}}, \ and\ \bibinfo {author} {\bibfnamefont {Liang}\ \bibnamefont
  {Fu}},\ }\href@noop {} {\enquote {\bibinfo {title} {Moiré quantum chemistry:
  charge transfer in transition metal dichalcogenide superlattices},}\ }
  (\bibinfo {year} {2019}),\ \Eprint {http://arxiv.org/abs/1910.14061}
  {arXiv:1910.14061 [cond-mat.str-el]} \BibitemShut {NoStop}%
\bibitem [{\citenamefont {Xia}\ \emph {et~al.}(2009{\natexlab{b}})\citenamefont
  {Xia}, \citenamefont {Qian}, \citenamefont {Hsieh}, \citenamefont {Wray},
  \citenamefont {Pal}, \citenamefont {Lin}, \citenamefont {Bansil},
  \citenamefont {Grauer}, \citenamefont {Hor}, \citenamefont {Cava},\ and\
  \citenamefont {Hasan}}]{Bi2Se31}%
  \BibitemOpen
  \bibfield  {author} {\bibinfo {author} {\bibfnamefont {Y.}~\bibnamefont
  {Xia}}, \bibinfo {author} {\bibfnamefont {D.}~\bibnamefont {Qian}}, \bibinfo
  {author} {\bibfnamefont {D.}~\bibnamefont {Hsieh}}, \bibinfo {author}
  {\bibfnamefont {L.}~\bibnamefont {Wray}}, \bibinfo {author} {\bibfnamefont
  {A.}~\bibnamefont {Pal}}, \bibinfo {author} {\bibfnamefont {H.}~\bibnamefont
  {Lin}}, \bibinfo {author} {\bibfnamefont {A.}~\bibnamefont {Bansil}},
  \bibinfo {author} {\bibfnamefont {D.}~\bibnamefont {Grauer}}, \bibinfo
  {author} {\bibfnamefont {Y.~S.}\ \bibnamefont {Hor}}, \bibinfo {author}
  {\bibfnamefont {R.~J.}\ \bibnamefont {Cava}}, \ and\ \bibinfo {author}
  {\bibfnamefont {M.~Z.}\ \bibnamefont {Hasan}},\ }\bibfield  {title} {\enquote
  {\bibinfo {title} {Observation of a large-gap topological-insulator class
  with a single dirac cone on the surface},}\ }\href {\doibase
  10.1038/nphys1274} {\bibfield  {journal} {\bibinfo  {journal} {Nature
  Physics}\ }\textbf {\bibinfo {volume} {5}},\ \bibinfo {pages} {398--402}
  (\bibinfo {year} {2009}{\natexlab{b}})}\BibitemShut {NoStop}%
\bibitem [{\citenamefont {Hsieh}\ \emph {et~al.}(2009)\citenamefont {Hsieh},
  \citenamefont {Xia}, \citenamefont {Qian}, \citenamefont {Wray},
  \citenamefont {Dil}, \citenamefont {Meier}, \citenamefont {Osterwalder},
  \citenamefont {Patthey}, \citenamefont {Checkelsky}, \citenamefont {Ong},
  \citenamefont {Fedorov}, \citenamefont {Lin}, \citenamefont {Bansil},
  \citenamefont {Grauer}, \citenamefont {Hor}, \citenamefont {Cava},\ and\
  \citenamefont {Hasan}}]{Bi2Se32}%
  \BibitemOpen
  \bibfield  {author} {\bibinfo {author} {\bibfnamefont {D.}~\bibnamefont
  {Hsieh}}, \bibinfo {author} {\bibfnamefont {Y.}~\bibnamefont {Xia}}, \bibinfo
  {author} {\bibfnamefont {D.}~\bibnamefont {Qian}}, \bibinfo {author}
  {\bibfnamefont {L.}~\bibnamefont {Wray}}, \bibinfo {author} {\bibfnamefont
  {J.~H.}\ \bibnamefont {Dil}}, \bibinfo {author} {\bibfnamefont
  {F.}~\bibnamefont {Meier}}, \bibinfo {author} {\bibfnamefont
  {J.}~\bibnamefont {Osterwalder}}, \bibinfo {author} {\bibfnamefont
  {L.}~\bibnamefont {Patthey}}, \bibinfo {author} {\bibfnamefont {J.~G.}\
  \bibnamefont {Checkelsky}}, \bibinfo {author} {\bibfnamefont {N.~P.}\
  \bibnamefont {Ong}}, \bibinfo {author} {\bibfnamefont {A.~V.}\ \bibnamefont
  {Fedorov}}, \bibinfo {author} {\bibfnamefont {H.}~\bibnamefont {Lin}},
  \bibinfo {author} {\bibfnamefont {A.}~\bibnamefont {Bansil}}, \bibinfo
  {author} {\bibfnamefont {D.}~\bibnamefont {Grauer}}, \bibinfo {author}
  {\bibfnamefont {Y.~S.}\ \bibnamefont {Hor}}, \bibinfo {author} {\bibfnamefont
  {R.~J.}\ \bibnamefont {Cava}}, \ and\ \bibinfo {author} {\bibfnamefont
  {M.~Z.}\ \bibnamefont {Hasan}},\ }\bibfield  {title} {\enquote {\bibinfo
  {title} {A tunable topological insulator in the spin helical dirac transport
  regime},}\ }\href {https://doi.org/10.1038/nature08234} {\bibfield  {journal}
  {\bibinfo  {journal} {Nature}\ }\textbf {\bibinfo {volume} {460}},\ \bibinfo
  {pages} {1101 EP --} (\bibinfo {year} {2009})}\BibitemShut {NoStop}%
\bibitem [{\citenamefont {Das~Sarma}\ and\ \citenamefont {Li}(2013)}]{Sarma}%
  \BibitemOpen
  \bibfield  {author} {\bibinfo {author} {\bibfnamefont {S.}~\bibnamefont
  {Das~Sarma}}\ and\ \bibinfo {author} {\bibfnamefont {Qiuzi}\ \bibnamefont
  {Li}},\ }\bibfield  {title} {\enquote {\bibinfo {title} {Many-body effects
  and possible superconductivity in the two-dimensional metallic surface states
  of three-dimensional topological insulators},}\ }\href {\doibase
  10.1103/PhysRevB.88.081404} {\bibfield  {journal} {\bibinfo  {journal} {Phys.
  Rev. B}\ }\textbf {\bibinfo {volume} {88}},\ \bibinfo {pages} {081404}
  (\bibinfo {year} {2013})}\BibitemShut {NoStop}%
\bibitem [{\citenamefont {Zhu}\ \emph {et~al.}(2012)\citenamefont {Zhu},
  \citenamefont {Santos}, \citenamefont {Howard}, \citenamefont {Sankar},
  \citenamefont {Chou}, \citenamefont {Chamon},\ and\ \citenamefont
  {El-Batanouny}}]{HeliumScattering}%
  \BibitemOpen
  \bibfield  {author} {\bibinfo {author} {\bibfnamefont {Xuetao}\ \bibnamefont
  {Zhu}}, \bibinfo {author} {\bibfnamefont {L.}~\bibnamefont {Santos}},
  \bibinfo {author} {\bibfnamefont {C.}~\bibnamefont {Howard}}, \bibinfo
  {author} {\bibfnamefont {R.}~\bibnamefont {Sankar}}, \bibinfo {author}
  {\bibfnamefont {F.~C.}\ \bibnamefont {Chou}}, \bibinfo {author}
  {\bibfnamefont {C.}~\bibnamefont {Chamon}}, \ and\ \bibinfo {author}
  {\bibfnamefont {M.}~\bibnamefont {El-Batanouny}},\ }\bibfield  {title}
  {\enquote {\bibinfo {title} {Electron-phonon coupling on the surface of the
  topological insulator {Bi$_2$Se$_3$} determined from surface-phonon
  dispersion measurements},}\ }\href {\doibase 10.1103/PhysRevLett.108.185501}
  {\bibfield  {journal} {\bibinfo  {journal} {Phys. Rev. Lett.}\ }\textbf
  {\bibinfo {volume} {108}},\ \bibinfo {pages} {185501} (\bibinfo {year}
  {2012})}\BibitemShut {NoStop}%
\bibitem [{\citenamefont {Hatch}\ \emph {et~al.}(2011)\citenamefont {Hatch},
  \citenamefont {Bianchi}, \citenamefont {Guan}, \citenamefont {Bao},
  \citenamefont {Mi}, \citenamefont {Iversen}, \citenamefont {Nilsson},
  \citenamefont {Hornek\ae{}r},\ and\ \citenamefont {Hofmann}}]{ARPES1}%
  \BibitemOpen
  \bibfield  {author} {\bibinfo {author} {\bibfnamefont {Richard~C.}\
  \bibnamefont {Hatch}}, \bibinfo {author} {\bibfnamefont {Marco}\ \bibnamefont
  {Bianchi}}, \bibinfo {author} {\bibfnamefont {Dandan}\ \bibnamefont {Guan}},
  \bibinfo {author} {\bibfnamefont {Shining}\ \bibnamefont {Bao}}, \bibinfo
  {author} {\bibfnamefont {Jianli}\ \bibnamefont {Mi}}, \bibinfo {author}
  {\bibfnamefont {Bo~Brummerstedt}\ \bibnamefont {Iversen}}, \bibinfo {author}
  {\bibfnamefont {Louis}\ \bibnamefont {Nilsson}}, \bibinfo {author}
  {\bibfnamefont {Liv}\ \bibnamefont {Hornek\ae{}r}}, \ and\ \bibinfo {author}
  {\bibfnamefont {Philip}\ \bibnamefont {Hofmann}},\ }\bibfield  {title}
  {\enquote {\bibinfo {title} {Stability of the {Bi$_2$Se$_3$}(111) topological
  state: Electron-phonon and electron-defect scattering},}\ }\href {\doibase
  10.1103/PhysRevB.83.241303} {\bibfield  {journal} {\bibinfo  {journal} {Phys.
  Rev. B}\ }\textbf {\bibinfo {volume} {83}},\ \bibinfo {pages} {241303}
  (\bibinfo {year} {2011})}\BibitemShut {NoStop}%
\bibitem [{\citenamefont {Pan}\ \emph {et~al.}(2012)\citenamefont {Pan},
  \citenamefont {Fedorov}, \citenamefont {Gardner}, \citenamefont {Lee},
  \citenamefont {Chu},\ and\ \citenamefont {Valla}}]{ARPES2}%
  \BibitemOpen
  \bibfield  {author} {\bibinfo {author} {\bibfnamefont {Z.-H.}\ \bibnamefont
  {Pan}}, \bibinfo {author} {\bibfnamefont {A.~V.}\ \bibnamefont {Fedorov}},
  \bibinfo {author} {\bibfnamefont {D.}~\bibnamefont {Gardner}}, \bibinfo
  {author} {\bibfnamefont {Y.~S.}\ \bibnamefont {Lee}}, \bibinfo {author}
  {\bibfnamefont {S.}~\bibnamefont {Chu}}, \ and\ \bibinfo {author}
  {\bibfnamefont {T.}~\bibnamefont {Valla}},\ }\bibfield  {title} {\enquote
  {\bibinfo {title} {Measurement of an exceptionally weak electron-phonon
  coupling on the surface of the topological insulator {Bi$_2$Se$_3$} using
  angle-resolved photoemission spectroscopy},}\ }\href {\doibase
  10.1103/PhysRevLett.108.187001} {\bibfield  {journal} {\bibinfo  {journal}
  {Phys. Rev. Lett.}\ }\textbf {\bibinfo {volume} {108}},\ \bibinfo {pages}
  {187001} (\bibinfo {year} {2012})}\BibitemShut {NoStop}%
\bibitem [{\citenamefont {Zhou}(2014)}]{SurfaceStateBandwidth}%
  \BibitemOpen
  \bibfield  {author} {\bibinfo {author} {\bibfnamefont {Wenwen}\ \bibnamefont
  {Zhou}},\ }\emph {\bibinfo {title} {STM probe on the surface electronic
  states of spin-orbit coupled materials}},\ \href@noop {} {Ph.D. thesis},\
  \bibinfo  {school} {Boston College} (\bibinfo {year} {2014})\BibitemShut
  {NoStop}%
\bibitem [{\citenamefont {Cano}\ \emph {et~al.}(2020)\citenamefont {Cano},
  \citenamefont {Fang}, \citenamefont {Pixley},\ and\ \citenamefont
  {Wilson}}]{Jennifer}%
  \BibitemOpen
  \bibfield  {author} {\bibinfo {author} {\bibfnamefont {Jennifer}\
  \bibnamefont {Cano}}, \bibinfo {author} {\bibfnamefont {Shiang}\ \bibnamefont
  {Fang}}, \bibinfo {author} {\bibfnamefont {J.~H.}\ \bibnamefont {Pixley}}, \
  and\ \bibinfo {author} {\bibfnamefont {Justin~H.}\ \bibnamefont {Wilson}},\
  }\href@noop {} {\enquote {\bibinfo {title} {A moiré superlattice on the
  surface of a topological insulator},}\ } (\bibinfo {year} {2020}),\ \Eprint
  {http://arxiv.org/abs/arXiv:2010.09726} {arXiv:2010.09726} \BibitemShut
  {NoStop}%
\bibitem [{\citenamefont {Bardeen}\ \emph
  {et~al.}(1957{\natexlab{a}})\citenamefont {Bardeen}, \citenamefont {Cooper},\
  and\ \citenamefont {Schrieffer}}]{BCS1}%
  \BibitemOpen
  \bibfield  {author} {\bibinfo {author} {\bibfnamefont {J.}~\bibnamefont
  {Bardeen}}, \bibinfo {author} {\bibfnamefont {L.~N.}\ \bibnamefont {Cooper}},
  \ and\ \bibinfo {author} {\bibfnamefont {J.~R.}\ \bibnamefont {Schrieffer}},\
  }\bibfield  {title} {\enquote {\bibinfo {title} {Microscopic theory of
  superconductivity},}\ }\href {\doibase 10.1103/PhysRev.106.162} {\bibfield
  {journal} {\bibinfo  {journal} {Phys. Rev.}\ }\textbf {\bibinfo {volume}
  {106}},\ \bibinfo {pages} {162--164} (\bibinfo {year}
  {1957}{\natexlab{a}})}\BibitemShut {NoStop}%
\bibitem [{\citenamefont {Bardeen}\ \emph
  {et~al.}(1957{\natexlab{b}})\citenamefont {Bardeen}, \citenamefont {Cooper},\
  and\ \citenamefont {Schrieffer}}]{BCS2}%
  \BibitemOpen
  \bibfield  {author} {\bibinfo {author} {\bibfnamefont {J.}~\bibnamefont
  {Bardeen}}, \bibinfo {author} {\bibfnamefont {L.~N.}\ \bibnamefont {Cooper}},
  \ and\ \bibinfo {author} {\bibfnamefont {J.~R.}\ \bibnamefont {Schrieffer}},\
  }\bibfield  {title} {\enquote {\bibinfo {title} {Theory of
  superconductivity},}\ }\href {\doibase 10.1103/PhysRev.108.1175} {\bibfield
  {journal} {\bibinfo  {journal} {Phys. Rev.}\ }\textbf {\bibinfo {volume}
  {108}},\ \bibinfo {pages} {1175--1204} (\bibinfo {year}
  {1957}{\natexlab{b}})}\BibitemShut {NoStop}%
\bibitem [{\citenamefont {Tang}(2015)}]{AndersonMorelFlat}%
  \BibitemOpen
  \bibfield  {author} {\bibinfo {author} {\bibfnamefont {Evelyn}\ \bibnamefont
  {Tang}},\ }\emph {\bibinfo {title} {Topological phases in narrow-band
  systems}},\ \href@noop {} {Ph.D. thesis},\ \bibinfo  {school} {Massachusetts
  Institute of Technology} (\bibinfo {year} {2015})\BibitemShut {NoStop}%
\bibitem [{\citenamefont {Heikkil{\"a}}\ \emph {et~al.}(2011)\citenamefont
  {Heikkil{\"a}}, \citenamefont {Kopnin},\ and\ \citenamefont
  {Volovik}}]{Grisha}%
  \BibitemOpen
  \bibfield  {author} {\bibinfo {author} {\bibfnamefont {T.~T.}\ \bibnamefont
  {Heikkil{\"a}}}, \bibinfo {author} {\bibfnamefont {N.~B.}\ \bibnamefont
  {Kopnin}}, \ and\ \bibinfo {author} {\bibfnamefont {G.~E.}\ \bibnamefont
  {Volovik}},\ }\bibfield  {title} {\enquote {\bibinfo {title} {Flat bands in
  topological media},}\ }\href {\doibase 10.1134/S0021364011150045} {\bibfield
  {journal} {\bibinfo  {journal} {JETP Letters}\ }\textbf {\bibinfo {volume}
  {94}},\ \bibinfo {pages} {233} (\bibinfo {year} {2011})}\BibitemShut
  {NoStop}%
\bibitem [{\citenamefont {Ojaj\"arvi}\ \emph {et~al.}(2018)\citenamefont
  {Ojaj\"arvi}, \citenamefont {Hyart}, \citenamefont {Silaev},\ and\
  \citenamefont {Heikkil\"a}}]{Tero}%
  \BibitemOpen
  \bibfield  {author} {\bibinfo {author} {\bibfnamefont {Risto}\ \bibnamefont
  {Ojaj\"arvi}}, \bibinfo {author} {\bibfnamefont {Timo}\ \bibnamefont
  {Hyart}}, \bibinfo {author} {\bibfnamefont {Mihail~A.}\ \bibnamefont
  {Silaev}}, \ and\ \bibinfo {author} {\bibfnamefont {Tero~T.}\ \bibnamefont
  {Heikkil\"a}},\ }\bibfield  {title} {\enquote {\bibinfo {title} {Competition
  of electron-phonon mediated superconductivity and stoner magnetism on a flat
  band},}\ }\href {\doibase 10.1103/PhysRevB.98.054515} {\bibfield  {journal}
  {\bibinfo  {journal} {Phys. Rev. B}\ }\textbf {\bibinfo {volume} {98}},\
  \bibinfo {pages} {054515} (\bibinfo {year} {2018})}\BibitemShut {NoStop}%
\bibitem [{\citenamefont {Peltonen}\ \emph {et~al.}(2018)\citenamefont
  {Peltonen}, \citenamefont {Ojaj\"arvi},\ and\ \citenamefont
  {Heikkil\"a}}]{Tero2}%
  \BibitemOpen
  \bibfield  {author} {\bibinfo {author} {\bibfnamefont {Teemu~J.}\
  \bibnamefont {Peltonen}}, \bibinfo {author} {\bibfnamefont {Risto}\
  \bibnamefont {Ojaj\"arvi}}, \ and\ \bibinfo {author} {\bibfnamefont
  {Tero~T.}\ \bibnamefont {Heikkil\"a}},\ }\bibfield  {title} {\enquote
  {\bibinfo {title} {Mean-field theory for superconductivity in twisted bilayer
  graphene},}\ }\href {\doibase 10.1103/PhysRevB.98.220504} {\bibfield
  {journal} {\bibinfo  {journal} {Phys. Rev. B}\ }\textbf {\bibinfo {volume}
  {98}},\ \bibinfo {pages} {220504} (\bibinfo {year} {2018})}\BibitemShut
  {NoStop}%
\end{thebibliography}%

\end{document}